\documentclass[preprints,review,accept,pdftex,moreauthors]{Definitions/mdpi} 

\firstpage{1} 
\pubvolume{1}
\issuenum{1}
\articlenumber{0}
\pubyear{2022}
\copyrightyear{2022}
\datereceived{} 
\dateaccepted{} 
\datepublished{} 
\hreflink{https://doi.org/} 

\newcommand{\be}{\begin{equation}}
\newcommand{\ee}{\end{equation}}
\newcommand{\ba}{\begin{eqnarray}}
\newcommand{\ea}{\end{eqnarray}}
\newcommand{\nn}{\nonumber}

\def\be{\begin{equation}}
\def\ee{\end{equation}}
\def\bea{\begin{eqnarray}}
\def\eea{\end{eqnarray}}
\def\eqi{\begin{equation}}
\def\eqf{\end{equation}}
\def\eqia{\begin{eqnarray}}
\def\eqfa{\end{eqnarray}}
\def\lcdm{$\Lambda$CDM }

\newcommand{\vp}{\varphi}
\newcommand{\mc}{\mathcal}
\renewcommand{\(}{\left(}
\renewcommand{\)}{\right)}

\pdfoutput=1

\Title{The Effective Fluid approach for Modified Gravity and its applications}

\TitleCitation{The Effective Fluid approach for Modified Gravity and its applications}

\Author{Savvas Nesseris $^{1}$\orcidA{}}

\AuthorNames{Savvas Nesseris}

\AuthorCitation{Nesseris, S.}

\address{%
$^{1}$ \quad Instituto de F\'isica Te\'orica UAM-CSIC, Universidad Auton\'oma de Madrid, Cantoblanco, 28049 Madrid, Spain; savvas.nesseris@csic.es}

\abstract{In this review we briefly summarize the so-called effective fluid approach, which is a compact framework that can be used to describe a plethora of different modified gravity models as general relativity (GR) and a dark energy (DE) fluid. This approach, which is complementary to the cosmological effective field theory, has several benefits as it allows for the easier inclusion of most modified gravity models into the state-of-the-art Boltzmann codes, that are typically hard-coded for GR and DE. Furthermore, it can also provide theoretical insights into their behavior, since in linear perturbation theory it is easy to derive physically motivated quantities such as the DE anisotropic stress or the DE sound speed. We also present some explicit applications of the effective fluid approach with $f(R)$, Horndeski and Scalar-Vector-Tensor models, namely how this approach can be used to easily solve the perturbation equations and incorporate the aforementioned modified gravity models into Boltzmann codes so as to obtain cosmological constraints using Monte Carlo analyses.}


\begin{document}

\section{Introduction}
At the end of the previous century statistically significant evidence from observations of Type Ia supernovae (SnIa) revealed that the Universe is currently undergoing a phase of accelerated expansion \cite{Riess:1998cb,Perlmutter:1998np}. This accelerated expansion is typically attributed to a cosmological constant $\Lambda$, which in addition to the standard Cold Dark Matter (CDM) scenario, can alleviate several deficiencies of the latter \cite{Kofman:1985fp}. Together these two components form the \lcdm model, which has been found to be in excellent agreement with recent cosmological measurements \cite{Aghanim:2018eyx,Abbott:2017wau}. Despite that, still the cosmological constant provides a problem for theoretical physics due to the large discrepancy between the predicted and observed values of $\Lambda$ \cite{Weinberg:1988cp,Carroll:2000fy}.

Since then, the \lcdm model has become the standard cosmological model as it provides the best description of the observations on cosmological scales \cite{Hinshaw:2012aka,Aghanim:2018eyx,Abbott:2017wau}. As a consequence, many alternative explanations have emerged and for the most part there are two main approaches. The first one, which is also more concretely based on high energy physics is the one where Dark Energy (DE) models \cite{Copeland:2006wr}, due to as yet unobserved scalar fields, dominate the energy budget of the Universe at late times, and if their mass is sufficiently light they also lead to an accelerated expansion \cite{Ratra:1987rm,ArmendarizPicon:2000dh}. The second approach is based on the assumption that covariant corrections to the Theory of General Relativity (GR) can alter gravity, usually dubbed modified gravity (MG), at sufficiently large scales \cite{Clifton:2011jh}. However, several cosmological probes in extra-galactic scales are in good agreement with GR \cite{Collett:2018gpf,LIGOScientific:2016lio}.

Overall, both approaches with either DE or MG models provide realistic and plausible explanations for the accelerated expansion of the Universe at late times. Furthermore, both kinds of models can also fit the cosmological observations at the background level equally well to the \lcdm model, as they can always go arbitrarily close to the cosmological constant. Thus, these models are in principle degenerate at the background level, despite laborious efforts to break the degeneracies with model independent approaches \cite{Nesseris:2010ep,Nesseris:2012tt}. Fortuitously, the recent discovery of gravitational waves by the joint LIGO/Virgo collaboration \cite{Abbott:2017oio} has allowed the community to rule out several MG models \cite{Creminelli:2017sry,Sakstein:2017xjx,Ezquiaga:2017ekz,Baker:2017hug,Amendola:2017orw,Crisostomi:2017pjs,Frusciante:2018jzw,Kase:2018aps,McManus:2016kxu,Lombriser:2015sxa}.

One of the main remaining classes of MG models is the so-called $f(R)$ models \cite{Sotiriou:2008rp,DeFelice:2010aj,Nojiri:2017ncd,Nojiri:2010wj}. Also in this case the background evolution of the Universe is degenerated with DE models with a redshift dependent equation of state $w(z)$ \cite{Multamaki:2005zs,delaCruzDombriz:2006fj,Pogosian:2007sw,Nesseris:2013fca}), albeit the linear theory perturbations are in general vastly different and have a particular time and scale dependence  \cite{Tsujikawa:2007gd}. This is particularly important as in general the DE perturbations have a definite effect on the growth-rate of matter perturbations and the so-called growth-index $\gamma$ \cite{Nesseris:2015fqa}. However, tat this point current data do not favor any particular model $f(R)$ model \cite{Luna:2018tot,Perez-Romero:2017njc}.

This discussion obviously makes it clear that the perturbations of MG models are of great importance and several approaches exist in the literature, e.g. see Refs.~\cite{Hu:2007nk,Hu:2007pj,Tsujikawa:2007gd,Kunz:2006ca,Pogosian:2007sw,Koivisto:2006xf,Koivisto:2006ai,delaCruzDombriz:2008cp,delaCruzDombriz:2006fj,Hu:2007pj,Starobinsky:2007hu,Bean:2006up,Song:2010rm,Pogosian:2010tj,Bean:2010zq,Caldwell:2007cw,Bertschinger:2008zb,Baker:2011jy,Silvestri:2013ne,Clifton:2018cef,Ishak:2018his}. However, even though many authors consider MG models, they still sometimes fix the background to that of the flat \lcdm model, as for example was done in Ref.~\cite{Zhao:2008bn}, but model the evolution of the Newtonian potentials via two functions $\mu(a,k)$ and $\gamma(a,k)$ which take into account possible deviations from GR. These  functions have been implemented in modified versions of the codes CAMB \cite{Lewis:1999bs} and CLASS \cite{Blas:2011rf}. Even though these parameterizations are only valid at late times, also new parameterizations which are valid at all times have appeared for MGCAMB, see Ref.~\cite{Hojjati:2011ix}, and MGCLASS, see Ref.~\cite{Sakr:2021ylx}, which is build to be compatible with parameter estimation codes. A definite flaw of the approach in early versions of MGCAMB was that the background expansion was fixed to that of the \lcdm model, even though it is known that for $f(R)$ models it is quite different, as is for example the case for the Hu-Sawicki model \cite{Hu:2007nk}.

A complimentary approach was carried out in Ref.~\cite{He:2012wq}, where the author studied perturbations of $f(R)$, which are degenerate to the \lcdm at the background level, by utilizing the full set of covariant cosmological perturbation equations and modifying  the publicly available code CAMB, in a new version called FRCAMB. Furthermore, an unreleased extended version of the aforementioned code with arbitrary background expansion rates was created by Ref.~\cite{Xu:2015usa}.

A totally different approach to modeling DE and MG models have been proposed in the form of the Effective Field Theory (EFT) \cite{Gubitosi:2012hu} and applied to cosmology in Ref. \cite{Hu:2013twa} in the form of a code called EFTCAMB. The main aadvantage of this approach is that it does not utilize any approximations, albeit the mapping of specific MG and DE models into the EFT formalism is somewhat complicated in most cases. Some of the aforementioned codes, namely MGCAMB and EFTCAMB, where used by the Planck Collaboration in Ref. \cite{Ade:2015rim} to derive cosmological constraints of the MG and DE models. Overall, no conclusive and statistically significant evidence for models beyond \lcdm was found.

Finally, another interesting approach was followed by Ref.~\cite{Battye:2015hza} where the authors proposed the so-called Equation of State (EOS) approach for perturbations which maps $f(R)$ models to a DE fluid at both the background and linear perturbation order \cite{Kunz:2006ca,Pogosian:2010tj}, see also \cite{Capozziello:2005mj,Capozziello:2006dj,Capozziello:2018ddp}. The EOS approach has been implemented in a modified version of the code CLASS \cite{Blas:2011rf} in Ref.~\cite{Battye:2017ysh}, albeit the problem with this approach is that the interpretation of the perturbation variables is not clear.

In this work we will also map the MG models as a DE fluid by utilizing the DE equation of state $w(a)$, the sound speed $c_s^2(a,k)$ and the anisotropic stress $\pi(a,k)$, as these variables are enough to describe any MG fluid at the background and linear order of perturbations \cite{Kunz:2012aw}. This has the advantage that it makes the comparison with popular DE models such as quintessence ($w(a)\geq -1$, $c_s^2=1$, $\pi(a,k)=0$) and K-essence ($w(a)$, $c_s^2(a)$, $\pi(a,k)=0$) straightforward. This is clearly important, as in general in MG models the anisotropic stress is non-zero  $\pi(a,k) \neq 0$, whereas in standard quintessence $\pi(a,k) = 0$, so that any statistically significant deviation of the anisotropic stress from zero, would be a smoking gun for MG models \cite{Saltas:2011xlz,Kunz:2012aw}.

In order to simplify the analysis of the perturbation equations, the quasi-static and sub-horizon approximations are frequently utilized. The former is based on the observation that in matter domination the Newtonian potentials are mostly constant, thus terms in the linearized Einstein equations with potentials with time derivatives can for the most part be safely neglected. The latter, is based on the observation that only perturbations with wavelengths shorter than the cosmological horizon are important. Some of the previous codes, i.e. FRCAMB, EFTCAMB, CLASS\_EOS\_FR did not in fact apply the sub-horizon approximation to the perturbation equations, albeit the quasi-static approximation has been studied extensively and implemented in MGCAMB \cite{delaCruzDombriz:2008cp,Sawicki:2015zya}. 

It has been argued, see for example Ref.~\cite{Sawicki:2015zya}, that the quasi-static approximation breaks down outside the DE sound-horizon $k\ll k_J$, where $k_J(z)\equiv \frac{H(z)}{(1+z) c_s}$ is the physical Jeans scale, rather than outside the cosmological horizon. However, in the aforementioned analysis the anisotropic stress was neglected and a constant DE $c_s^2$ was utilized, both assumptions being unrealistic in general. 

This approach allows us in general to discriminate between traditional DE and MG models, as they both have vastly different predictions for the equation of state $w(a)$, the sound speed $c_s^2(a,k)$, and the anisotropic stress $\pi(a,k)$. The last two quantities are in particular important as they in general may leave observable traces in the large scale structure (LSS), the Cosmic Microwave Background radiation (CMB) and Galaxy Counts (GC) \cite{Tsujikawa:2007gd,Cardona:2014iba}. Furthermore, an important aspect of the anisotropic stress is that in general it can stabilize the growth of matter perturbations in cases where that would be not possible \cite{Saltas:2011xlz,Koivisto:2005mm,Mota:2007sz,Cardona:2014iba}, while the sound speed affects the clustering of matter perturbations \cite{Hu:1998kj,dePutter:2010vy,Batista:2017lwf}. Both of these effects are crucial, as they can be used to break the parameter degeneracies between the models \cite{Lewis:2002ah,Tegmark:2003ud}.

Even though the \lcdm model seems to be in good overall agreement with observations \cite{Aghanim:2018eyx,Abbott:2017wau}, this might easily change when forthcoming galaxy surveys like Euclid, DESI and stage IV CMB experiments arrive. Furthermore, there also seem to remain some more issues with cosmological data such as direct Hubble constant measurements, weak lensing data, and cluster counts, where different aspects of DE models or MG models could be important \cite{Ade:2015xua,Zhao:2017cud,Heavens:2017hkr,Freedman:2017yms,Renk:2017rzu,Nunes:2018xbm,Lin:2018nxe,Cardona:2014iba,Benetti:2018zhv,Sakr:2018new}, thus the effective fluid approach would be particularly useful.

This review is organized as follows: in Sec.~\ref{sec:theory} we present the theoretical framework of the effective fluid approach and its application to  $f(R)$, Horndeski and Scalar-Vector-Tensor models. Then, in Sec.~\ref{sec:apps} we present several concrete applications of our approach, namely designer Horndeski models, the numerical solutions of the perturbation equations and the necessary modifications to Boltzmann codes so that comparison with the CMB data and Monte Carlo analyses can be made. Finally, in Sec.~\ref{sec:conc} we summarize our the effective fluid approach and present our conclusions.

\section{Theoretical framework\label{sec:theory}}
Here we now describe the theoretical framework necessary to illustrate the Effective Fluid approach, especially related to linear order of perturbation theory. On large scales the Universe is homogeneous and isotropic, thus it can be described at the background level by a Friedmann–Lema\^itre–Robertson–Walker (FLRW) metric. In order to describe the large scale structure of the Universe, we need to consider the perturbed FLRW metric, which in the conformal Newtonian gauge is given by:
\be
ds^2=a(\tau)^2\left[-(1+2\Psi(\vec{x},\tau))d\tau^2+(1-2\Phi(\vec{x},\tau))d\vec{x}^2\right],
\label{eq:FRWpert}
\ee
given in terms of the conformal time $\tau$ defined via $d\tau=dt/a(t)$ and we also follow the notation of Ref.~\cite{Ma:1995ey}.\footnote{In this review our conventions are: (-+++) for the metric signature, the Riemann and Ricci tensors are given by $V_{b;cd}-V_{b;dc}=V_a R^a_{bcd}$ and $R_{ab}=R^s_{asb}$. The Einstein equations are $G_{\mu\nu}=+\kappa T_{\mu\nu}$ for $\kappa=\frac{8\pi G_\mathrm{N}}{c^4}$ and $G_\mathrm{N}$ is the bare Newton's constant, while in what follows we set the speed of light $c=1$.}

On large cosmological scales, where the average density of the matter particle species is very low with respect to terrestrial ones, namely on the order of $\rho\sim \rho_\mathrm{cr}=1.8788\times 10^{-26}\,h^2\,\mathrm{kg}\,\mathrm{m}^{-3}$, this means we can assume the matter species can be described as ideal fluids with an energy momentum tensor
\be
T^\mu_{\nu}=P\delta^\mu_{\nu}+(\rho+P)U^\mu U_\nu,\label{eq:enten}
\ee
where $\rho$, $P$ are the fluid density and pressure, while $U^\mu=\frac{dx^\mu}{\sqrt{-ds^2}}$ is its velocity four-vector given to first order by $U^\mu=\frac{1}{a(\tau)}\left(1-\Psi,\vec{u}\right)$, which satisfies $U^\mu U_\mu=-1$ and we have defined $\vec{u}=\dot{\vec{x}}$ and $\dot{f}\equiv\frac{df}{d\tau}$.

Then, the elements of the energy momentum tensor are given to linear order of perturbations by:
\bea
T^0_0&=&-(\bar{\rho}+\delta \rho), \label{eq:effectTmn1}\\
T^0_i&=&(\bar{\rho}+\bar{P})u_i,\\
T^i_j&=& (\bar{P}+\delta P)\delta^i_j+\Sigma^i_j, \label{eq:effectTmn}
\eea
where $\bar{\rho},\bar{P}$ are background quantities, functions of time only. On the other hand, the perturbations of the fluid's density and pressure are given by $\delta \rho, \delta P$ and are functions of $(\vec{x},\tau)$. Finally, $\Sigma^i_j\equiv T^i_j-\delta^i_j T^k_k/3$ is an anisotropic stress tensor.

In the context of GR we find that the perturbed Einstein equations are given in the conformal Newtonian gauge by \cite{Ma:1995ey}:
\be
k^2\Phi+3\frac{\dot{a}}{a}\left(\dot{\Phi}+\frac{\dot{a}}{a}\Psi\right) = 4 \pi G_\mathrm{N} a^2 \delta T^0_0, \label{eq:phiprimeeq}
\ee
\be
k^2\left(\dot{\Phi}+\frac{\dot{a}}{a}\Psi\right) = 4 \pi G_\mathrm{N} a^2 (\bar{\rho}+\bar{P})\theta,\label{eq:phiprimeeq1}
\ee
\be
\ddot{\Phi}+\frac{\dot{a}}{a}(\dot{\Psi}+2\dot{\Phi})+\left(2\frac{\ddot{a}}{a}-
 \frac{\dot{a}^2}{a^2}\right)\Psi+\frac{k^2}{3}(\Phi-\Psi)
=\frac{4\pi}{3}G_\mathrm{N} a^2\delta T^i_i,
\ee
\be
k^2(\Phi-\Psi) = 12\pi G_\mathrm{N} a^2 (\bar{\rho}+\bar{P})\sigma \label{eq:anisoeq},
\ee
where the velocity is given by $\theta\equiv ik^ju_j$ and the anisotropic stress by $(\bar{\rho}+\bar{P})\sigma\equiv-(\hat{k}_i\hat{k}_j-\frac13 \delta_{ij})\Sigma^{ij}$, which as mentioned is related to the traceless part of the energy momentum tensor via $\Sigma^i_j\equiv T^i_j-\delta^i_j T^k_k/3$ and where $k^{j}$ and $\hat{k}^{j}$ are the wavenumbers and unit vectors in Fourier/$k$-space of the perturbations.

To find the evolution equations for the perturbation variables we use the energy-momentum conservation $T^{\mu\nu}{}_{;\nu}=0$  conservation given by the Bianchi identities in GR as:
\bea
\dot{\delta} &=& -(1+w)(\theta-3\dot{\Phi})-3\frac{\dot{a}}{a}\left(c_s^2-w\right)\delta, \label{eq:cons1}\\
\dot{\theta} &=&  -\frac{\dot{a}}{a}(1-3w)\theta-\frac{\dot{w}}{1+w}\theta+\frac{c_s^2}{1+w}k^2\delta-k^2\sigma+k^2\Psi,
\label{eq:cons2}
\eea
where we have defined the rest-frame sound speed of the fluid $c_s^2\equiv\frac{\delta P}{\delta \rho}$ and its equation of state parameter $w\equiv\frac{\bar{P}}{\bar{\rho}}$. After eliminating $\theta$ from Eqs.~(\ref{eq:cons1}) and (\ref{eq:cons2}) we find a second order equation for $\delta$ \cite{Cardona:2014iba}:
\bea
\ddot{\delta}+(\ldots) \dot{\delta}+(\ldots) \delta &=& - k^2\left((1+w)\Psi+c_s^2\delta-(1+w)\sigma\right)+\ldots \nn \\
&=& -k^2\left((1+w)\Psi+c_s^2\delta-\frac23\pi\right)+\ldots,
\eea
where the dots $(\ldots)$ stand for a complicated expression, while the anisotropic stress of the fluid is defined as $\pi\equiv\frac32(1+w)\sigma$. As can be seen, the $k^2$ term behaves as a source driving the perturbations of the fluid, however as the potential scales as $\Psi\sim1/k^2$ in relevant scales (due to the Poisson equation), the dominant terms are only the sound speed and the anisotropic stress. Thus, we can define a parameter, namely an effective sound speed which controls the stability of the perturbations, as \cite{Cardona:2014iba}:
\be
c_\mathrm{s,eff}^2 =c_s^2-\frac23\pi/\delta.\label{eq:cs2eff}
\ee
Clearly, $c_\mathrm{s,eff}^2$ not only characterizes the propagation of perturbations, it also defines the clustering properties of the fluid on sub-horizon scales, see Ref.~\cite{Cardona:2014iba}. In general, the sound speed $c_s^2$ can be both time and scale dependent, i.e., $c_s^2=c_s^2(\tau,k)$, e.g. as noted in Ref.~\cite{Amendola:2015ksp}, the sound speed in small scales for a scalar field $\phi$ (in the conformal Newtonian gauge)  is given by $c_\mathrm{s,\phi}^2\simeq\frac{k^2}{4 a^2 m_\phi^2}$, where $m_\phi$ is the mass of the scalar field. 

However, the sound speed is equal to unity only in the scalar field's rest-frame. see for example Chapter 11.2 of Ref.~\cite{Amendola:2015ksp}. In $f(R)$ theories, as they in fact can be viewed as a non-minimally coupled scalar field in the Einstein frame, see for example Ref.~\cite{Mukhanov:1990me, Sawicki:2015zya}, the sound speed is also scale dependent, when we are not in the rest frame of the equivalent DE fluid.

In the end, it is common to use the scalar velocity perturbation $V\equiv i k_jT^j_0/\rho=(1+w)\theta$, instead of the fluid velocity $\theta$ as the former can remain finite when the equation of state $-1$, see for example Ref.~\cite{Sapone:2009mb}. Then, Eqs.~\eqref{eq:cons1}-\eqref{eq:cons2} become

\bea
\delta' &=& 3(1+w) \Phi'-\frac{V}{a^2 H}-\frac{3}{a}\left(\frac{\delta P}{\bar{\rho}}-w\delta\right),
\label{Eq:evolution-delta} \\
V' &=& -(1-3w)\frac{V}{a}+\frac{k^2}{a^2 H}\frac{\delta P}{\bar{\rho}} +(1+w)\frac{k^2}{a^2 H} \Psi -\frac23 \frac{k^2}{a^2 H} \pi,
\label{Eq:evolution-V}
\eea
where a prime $'$ means a derivative with respect to the scale factor $a$, while $H(t)=\frac{da/dt}{a}$ is the cosmic-time Hubble parameter.

\subsection{$f(R)$ models \label{subsection:fr-efa}}
The simplest application of the effective fluid approach in theories beyond GR, is of course in the context of $f(R)$ models. These can of course be studied directly, as was done in Ref.~\cite{Tsujikawa:2007gd} or as an the effective DE fluid \cite{Battye:2015hza}. Specifically, the modified Einstein-Hilbert action is given by:
\be
S=\int d^{4}x\sqrt{-g}\left[  \frac{1}{2\kappa}f\left(  R\right)
+\mathcal{L}_\mathrm{m}\right],  \label{eq:action1}%
\ee
where $\mathcal{L}_\mathrm{m}$ is the matter Lagrangian and
$\kappa\equiv8\pi G_\mathrm{N}$. Varying the action of Eq.~\eqref{eq:action1} with respect to the metric, we obtain the field equations  \cite{Tsujikawa:2007gd}:
\be
F G_{\mu\nu}-\frac12(f(R)-R~F) g_{\mu\nu}+\left(g_{\mu\nu}\Box-\nabla_\mu\nabla_\nu\right)F =\kappa\,T_{\mu\nu}^{(m)},
\label{eq:EE}
\ee
where we have defined $F\equiv f'(R)$, $G_{\mu\nu}$ is the usual Einstein tensor and $T_{\mu\nu}^{(m)}$ is the energy-momentum tensor of the matter fields. 

Bu moving all the modified gravity contributions to the right hand side, we can rewrite the field equations as the Einstein equations being equal to the sum of the energy momentum tensors of the matter fields and that of an effective DE fluid \cite{Pogosian:2010tj}:
\bea
G_{\mu\nu}&=&\kappa\left(T_{\mu\nu}^{(m)}+T_{\mu\nu}^{(DE)}\right),
\label{eq:effEqs}
\eea
where we have defined
\be
\kappa T_{\mu\nu}^{(DE)}\equiv(1-F)G_{\mu\nu}+\frac12(f(R)-R~F) g_{\mu\nu} -\left(g_{\mu\nu}\Box-\nabla_\mu\nabla_\nu\right)F.
\label{eq:effTmn}
\ee
As $f(R)$ theories are also diffeomorphism invariant, one can show that the effective energy momentum tensor given by Eq.~\eqref{eq:effTmn} also satisfies a conservation equation:
\be
\nabla^\mu T_{\mu\nu}^{(DE)}=0.
\ee
Writing the theory in this way implies that the Friedmann equations are the same as in GR \cite{Ma:1995ey}:
\bea
\mathcal{H}^2&=&\frac{\kappa}{3}a^2 \left(\bar{\rho}_\mathrm{m}+\bar{\rho}_\mathrm{DE}\right), \\
\dot{\mathcal{H}}&=&-\frac{\kappa}{6}a^2 \big[\left(\bar{\rho}_\mathrm{m}+3\bar{P}_\mathrm{m}\right)+\left(\bar{\rho}_\mathrm{DE}+3\bar{P}_\mathrm{DE}\right)\big],
\eea
albeit with the addition of an effective DE term on the right hand side described by a density and pressure:
\bea
\kappa \bar{P}_\mathrm{DE}&=&\frac{f}2-\mathcal{H}^2/a^2-2F\mathcal{H}^2/a^2+\mathcal{H}\dot{F}/a^2 -\dot{\mathcal{H}}/a^2-F\dot{\mathcal{H}}/a^2+\ddot{F}/a^2,\label{eq:effpr}\\
\kappa \bar{\rho}_\mathrm{DE}&=&-\frac{f}2+3\mathcal{H}^2/a^2-3\mathcal{H}\dot{F}/a^2+3F\dot{\mathcal{H}}/a^2,\label{eq:effden}
\eea
where $\mathcal{H}=\frac{\dot{a}}{a}$ is the conformal Hubble parameter.
From Eqs.~(\ref{eq:effpr}) and (\ref{eq:effden}) we can define an effective DE equation of state for the $f(R)$ models as:
\be
w_\mathrm{DE}=\frac{-a^2 f+2\left((1+2F)\mathcal{H}^2-\mathcal{H}\dot{F}+(2+F)\dot{\mathcal{H}}-\ddot{F}\right)}{a^2 f-6(\mathcal{H}^2-\mathcal{H}\dot{F}+F\dot{\mathcal{H}})}\label{eq:wde},
\ee
which agrees with the one given in  Ref.~\cite{Tsujikawa:2007gd}.

Using the effective energy momentum tensor of Eq.~(\ref{eq:effTmn}) we can define the effective DE pressure, density and velocity perturbations as:
\bea
\frac{\delta P_\mathrm{DE}}{\bar{\rho}_\mathrm{DE}}&=&(...)\delta R+(...)\dot{\delta R}+(...)\ddot{\delta R}+(...)\Psi +(...)\dot{\Psi}+(...)\Phi+(...)\dot{\Phi},\label{eq:effdenp00}\\
\delta_\mathrm{DE}&=&(...)\delta R+(...)\dot{\delta R}+(...)\Psi+(...)\Phi +(...)\dot{\Phi},\label{eq:effprp0} \\
V_\mathrm{DE}&\equiv&(1+w_\mathrm{DE})\theta_\mathrm{DE} \nn \\
 &=&(...)\delta R+(...)\dot{\delta R}+(...)\Psi+(...)\Phi +(...)\dot{\Phi},\label{eq:efftheta0}
\eea
while the difference of the two Newtonian potentials $\Phi$ and $\Psi$ is given by
\be
\Phi-\Psi=\frac{F_{,R}}{F} \delta R.\label{eq:defaniso}
\ee
Thus, the anisotropic stress is given by \cite{Ma:1995ey}
\bea
\bar{\rho}_\mathrm{DE} \pi_\mathrm{DE}&=&-\frac32 (\hat{k}_i\hat{k}_j-\frac13 \delta_{ij})\Sigma^{ij} \nn \\
&=&\frac{1}{\kappa}\frac{k^2}{a^2}\left(F_{,R} \delta R+(1-F)(\Phi-\Psi)\right).
\label{Eq:DE-rho-anisotropic-stress}
\eea

\subsubsection{The quasi-static and sub-horizon approximations}
As can been seen, the expressions for the DE perturbations given by Eqs.~\eqref{eq:effdenp00}-\eqref{Eq:DE-rho-anisotropic-stress} are somewhat cumbersome and can be significantly simplified, without much loss of accuracy, by using the sub-horizon and quasi-static approximations. The former implies that only the modes deep in the Hubble radius $(k^2\gg a^2 H^2)$ are important, while the with the latter we neglect terms with time derivatives. As an example we find that the perturbation of the Ricci scalar is
\bea
\delta R&=&-\frac{12  (\mathcal{H}^2+\dot{\mathcal{H}})}{a^2}\Psi-\frac{4 k^2}{a^2}\Phi+\frac{2 k^2 }{a^2}\Psi -\frac{18 \mathcal{H} }{a^2}\dot{\Phi}-\frac{6 \mathcal{H} }{a^2}\dot{\Psi}-\frac{6 \ddot{\Phi}}{a^2},\label{eq:ricciexact}\\
&\simeq & -\frac{4 k^2}{a^2}\Phi+\frac{2 k^2}{a^2}\Psi,\label{eq:ricciapp}
\eea
where in the last line we applied the two approximations. Then from the perturbed Einstein equations we find the modified Poisson equations \cite{Tsujikawa:2007gd}:
\bea
\Psi &=& -4\pi G_\mathrm{N} \frac{a^2}{k^2} \frac{G_\mathrm{eff}}{G_\mathrm{N}} \bar{\rho}_\mathrm{m} \delta_\mathrm{m}, \label{eq:pot1} \\
\Phi &=& -4\pi G_\mathrm{N} \frac{a^2}{k^2} Q_\mathrm{eff} \bar{\rho}_\mathrm{m} \delta_\mathrm{m},\label{eq:pot2}
\eea
where $G_\mathrm{eff}$ and $Q_\mathrm{eff}$ are both equal to unity in GR and are given by \cite{Tsujikawa:2007gd}:
\bea
G_\mathrm{eff}/G_\mathrm{N}&=& \frac{1}{F} \frac{1+4\frac{k^2}{a^2}\frac{F_{,R}}{F}}{1+3\frac{k^2}{a^2}\frac{F_{,R}}{F}}, \label{eq:pot11} \\
Q_\mathrm{eff}&=& \frac{1}{F}  \frac{1+2\frac{k^2}{a^2}\frac{F_{,R}}{F}}{1+3\frac{k^2}{a^2}\frac{F_{,R}}{F}},\label{eq:pot21}
\eea
where we have set $F=\frac{d f(R)}{dR}$, $F_{,R}=\frac{d^2 f(R)}{dR^2}$. Alternatively we can also write the Poisson equation for $\Phi$ in the effective fluid approach, where we can to introduce the DE density $\rho_\mathrm{DE}$ and get:
\bea
-\frac{k^2}{a^2}\Phi&=&4 \pi G_\mathrm{N}\left( \bar{\rho}_\mathrm{m} \delta_\mathrm{m}+\bar{\rho}_\mathrm{DE}\delta_\mathrm{DE}\right)\nn \\
&=& 4 \pi G_\mathrm{N} Q_\mathrm{eff} \bar{\rho}_\mathrm{m} \delta_\mathrm{m},
\eea
which implies that 
\be
\bar{\rho}_\mathrm{m} \delta_\mathrm{m}=\frac{1}{Q_\mathrm{eff}-1}\bar{\rho}_\mathrm{DE}\delta_\mathrm{DE}.\label{eq:rmdm}\ee
which allows us to determine the evolution of the DE density perturbation directly.

With these approximations we can also directly derive a second order differential equation, when ignoring neutrinos, for the time evolution of the matter density contrast \cite{Tsujikawa:2007gd}:
\be
\delta_\mathrm{m}''(a)+\left(\frac{3}{a}+\frac{H'(a)}{H(a)}\right)\delta_\mathrm{m}'(a)-\frac32 \frac{\Omega_\mathrm{m0} G_\mathrm{eff}/G_\mathrm{N}}{a^5 H(a)^2/H_0^2}\delta_\mathrm{m}(a)=0, \label{eq:Geffode}
\ee
where derivatives with respect to the scale factor $a$ are denoted by a prime $'$. Finally, we can also define the DE anisotropic parameters
\bea
\eta &\equiv& \frac{\Psi-\Phi}{\Phi}\simeq \frac{2\frac{k^2}{a^2}\frac{F_{,R}}{F}}{1+2\frac{k^2}{a^2}\frac{F_{,R}}{F}}, \label{eq:an1} \\
\gamma &\equiv& \frac{\Phi}{\Psi}\simeq\frac{1+2\frac{k^2}{a^2}\frac{F_{,R}}{F}}{1+4\frac{k^2}{a^2}\frac{F_{,R}}{F}}. \label{eq:an2}
\eea

Applying the sub-horizon and quasi-static approximations, and using the Poisson equations, we can estimate the effective density, pressure and velocity perturbations of the DE fluid as:
\bea
\frac{\delta P_\mathrm{DE}}{\bar{\rho}_\mathrm{DE}}&\simeq&\frac{1}{3F}\frac{2\frac{k^2}{a^2}\frac{F_{,R}}{F}+3(1+5\frac{k^2}{a^2}\frac{F_{,R}}{F})\ddot{F}k^{-2}}{1+3\frac{k^2}{a^2}\frac{F_{,R}}{F}}\frac{\bar{\rho}_\mathrm{m}}{\bar{\rho}_\mathrm{DE}} \delta_\mathrm{m},\label{eq:effpres} \\
\delta_\mathrm{DE}&\simeq&\frac{1}{F}\frac{1-F+\frac{k^2}{a^2}(2-3F)\frac{F_{,R}}{F}}{1+3\frac{k^2}{a^2}\frac{F_{,R}}{F}}\frac{\bar{\rho}_\mathrm{m}}{\bar{\rho}_\mathrm{DE}} \delta_\mathrm{m},\label{eq:effder} \\
V_\mathrm{DE}&\equiv& (1+w_\mathrm{DE})\theta_\mathrm{DE}\nn \\ &\simeq&\frac{\dot{F}}{2F}\frac{1+6\frac{k^2}{a^2}\frac{F_{,R}}{F}}{1+3\frac{k^2}{a^2}\frac{F_{,R}}{F}}\frac{\bar{\rho}_\mathrm{m}}{\bar{\rho}_\mathrm{DE}} \delta_\mathrm{m},\label{eq:efftheta}
\eea
while the DE anisotropic stress parameter $\pi_\mathrm{DE}$ is given by
\bea
\pi_\mathrm{DE}&=& \frac{\frac{k^2}{a^2} (\Phi-\Psi)}{\kappa~ \bar{\rho}_\mathrm{DE}}\nn\\
&\simeq& \frac{1}{F}\frac{\frac{k^2}{a^2}\frac{F_{,R}}{F}}{1+3\frac{k^2}{a^2}\frac{F_{,R}}{F}}\frac{\bar{\rho}_\mathrm{m}}{\bar{\rho}_\mathrm{DE}} \delta_\mathrm{m} \nn\\
&\simeq&\frac{\frac{k^2}{a^2}\frac{F_{,R}}{F}}{1-F+\frac{k^2}{a^2}(2-3F)\frac{F_{,R}}{F}}\delta_\mathrm{DE}.\label{eq:effpi}
\eea
As can be seen, the DE anisotropic stress can also be written i nthe more general form as
\bea
\pi_\mathrm{DE}(a)&=& \frac{\frac{k^2}{a^2}f_1(a)}{1+\frac{k^2}{a^2}f_2(a)}\delta_\mathrm{DE}(a),
\eea
where we have defined the functions $f_1(a)=\frac{F_{,R}}{F (1-F)}$ and $f_2(a)=\frac{(2-3F)F_{,R}}{F (1-F)}$, which are reminiscent of Model 2 in Ref.~\cite{Cardona:2014iba}. 

Now using Eqs.~(\ref{eq:effpres}) and (\ref{eq:effder}), we find that effective DE sound speed of the fluid is given at this level of the approximation by
\be
c_\mathrm{s,DE}^2\simeq\frac13 \frac{2\frac{k^2}{a^2}\frac{F_{,R}}{F}+3(1+5\frac{k^2}{a^2}\frac{F_{,R}}{F})\ddot{F}k^{-2}}{1-F+\frac{k^2}{a^2}(2-3F)\frac{F_{,R}}{F}},
\label{eq:cs2de}
\ee
and the DE effective sound speed is
\bea
c_\mathrm{s,eff}^2 &\equiv& c_\mathrm{s,DE}^2-\frac23\pi_\mathrm{DE}/\delta_\mathrm{DE}\nn \\
&\simeq& \frac{(1+5\frac{k^2}{a^2}\frac{F_{,R}}{F})\ddot{F}k^{-2}}{1-F+\frac{k^2}{a^2}(2-3F)\frac{F_{,R}}{F}}.\label{eq:cs21}
\eea
As can be seen from the previous expressions, for the \lcdm model ($f(R)=R-2\Lambda$), we have $F=1$ and $F_{,R}=0$ implying $w_\mathrm{DE}=-1$ and $(\delta P_\mathrm{DE},\delta \rho_\mathrm{DE},\pi_\mathrm{DE})=(0,0,0)$ as it should. 

In the case of $f(R)$ models, such as the Hu \& Sawicki (HS, hereafter), where the DE equation of state $w_\mathrm{DE}$ crosses $w_\mathrm{DE}(a)=-1$, one would expect singularities to appear because of the $1+w$ term in the denominator in Eq.~(\ref{eq:cons2}) \cite{Nesseris:2006er}. However, we can absorb the $1+w$ term by introducing $V_\mathrm{DE}=(1+w_\mathrm{DE})\theta_\mathrm{DE}$, as this combination remains finite for well-behaved $f(R)$ models, as seen by inspecting Eq.~(\ref{eq:efftheta}). 

\begin{figure}[!t]
\centering
\includegraphics[width = 0.85\textwidth]{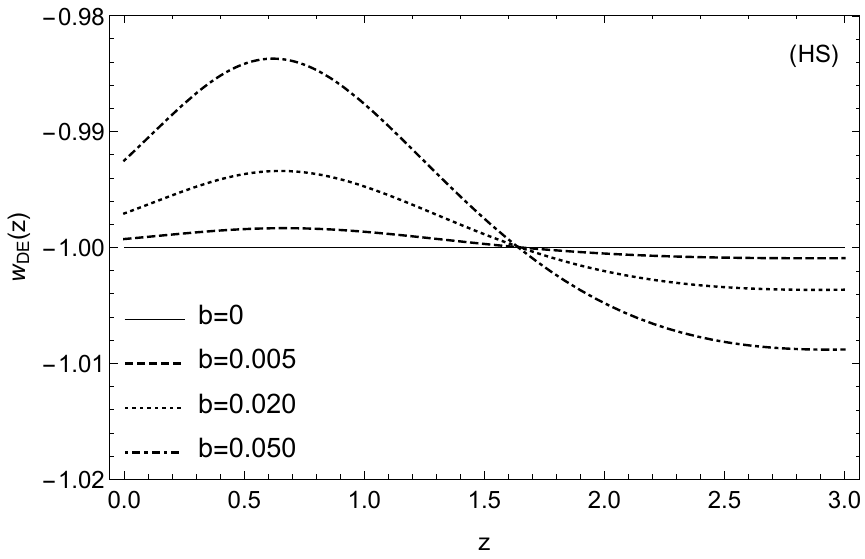}
\caption{The evolution of the effective DE equation of state $w_\mathrm{DE}(z)$ for the HS model for $\Omega_\mathrm{m0}=0.3$, $n=1$ and various values of the $b$ parameter which controls the deviations from the \lcdm model (see Ref.~\cite{Arjona:2018jhh}), with $b \in [0,0.05]$. The equation of state crosses $w_\mathrm{DE}=-1$ at $z\sim1.65$, while at early times we have $1+w_\mathrm{DE}<0$ thus violating the SEC. Image from Ref.~\cite{Arjona:2018jhh}.}
\label{fig:wdeHS}
\end{figure}

As a final remark, it should be noted that the effect DE fluid described here and in Eq. \eqref{eq:effTmn}, in fact violates the energy conditions of GR \cite{Wald:1984rg}, which can be expressed via the DE density and pressure:
\bea
{\bf NEC} &\Longrightarrow & \bar{\rho}_\mathrm{DE}+\bar{P}_\mathrm{DE} \ge 0, \nn \\
{\bf WEC} &\Longrightarrow &\bar{\rho}_\mathrm{DE}\ge 0 \hspace{3mm}\text{and}\hspace{3mm} \bar{\rho}_\mathrm{DE}+\bar{P}_\mathrm{DE} \ge 0, \nn \\
{\bf DEC} &\Longrightarrow & \bar{\rho}_\mathrm{DE}\ge 0 \hspace{3mm}\text{and}\hspace{3mm} \bar{\rho}_\mathrm{DE} \ge \left|\bar{P}_\mathrm{DE}\right|, \nn \\
{\bf SEC} &\Longrightarrow & \bar{\rho}_\mathrm{DE}+3\bar{P}_\mathrm{DE} \ge 0 \hspace{3mm}\text{and}\hspace{3mm} \bar{\rho}_\mathrm{DE}+\bar{P}_\mathrm{DE} \ge 0, \nn
\eea
where NEC, WEC, DEC and SEC stand for the null, weak, dominant and strong energy conditions. As the inequality  $\bar{\rho}_\mathrm{DE} \ge 0$ still holds, then the NEC, WEC and DEC conditions can be mapped equivalently to the constraint $w_\mathrm{DE} \ge -1$. However, as can be seen in Fig.~\ref{fig:wdeHS} in the case of the HS model the NEC, WEC and DEC are violated for redshifts $z \gtrsim 1.65$.

\subsection{Horndeski models}
The most general Lorentz-invariant extension of GR in four dimensions with a non-minimally coupled scalar field and second order equations of motion is the so-called Horndeski theory \cite{Horndeski:1974wa}. This theory contains several free functions that, in appropriate limits, reduce to several well-known DE and MG models. However, the recent observation of a binary neutron star and its accompanying optical counterpart, has produced an amazingly tight constraint on the speed of propagation of gravitational waves (GWs) \cite{LIGOScientific:2017zic}
\bea
-3 \cdot 10^{-15} \le c_g/c-1 \le 7 \cdot 10^{-16}.
\eea
This constraint implies that the functional forms of two of the free Horndeski functions were then limited to be \cite{Ezquiaga:2017ekz}
\bea
G_{4X}\approx 0, \hspace{2mm} G_5 \approx \text{const.},
\eea
as seen from the formula for the GW speed of propagation\cite{Kobayashi:2011nu}
\bea
c^2_T=\frac{G_4-XG_{5\phi}-XG_{5X}\ddot{\phi}}{G_4-2XG_{4X}-X\left(G_{5X}\dot{\phi}H-G_{5\phi}\right)}.
\eea
Thus, in the case of the effective fluid approach, we will only focus on the remaining parts of the Horndeski Lagrangian, in particular
\be
S[g_{\mu \nu}, \phi] = \int d^{4}x\sqrt{-g}\left[\sum^{4}_{i=2} \mathcal{L}_i\left[g_{\mu \nu},\phi\right] + \mathcal{L}_\mathrm{m} \right],
\label{eq:action11}
\ee
where we have defined
\bea
\mathcal{L}_2&=& G_2\left(\phi,X\right) \equiv K\left(\phi,X\right),\\
\mathcal{L}_3&=&-G_3\left(\phi,X\right)\Box \phi,\\
\mathcal{L}_4&=&G_4\left(\phi\right) R,
\eea
and $\phi$ is a scalar field, $X \equiv -\frac{1}{2}\partial_{\mu}\phi\partial^{\mu}\phi$ is its kinetic term, and $\Box \phi \equiv g^{\mu \nu}\nabla_{\mu}\nabla_{\nu}\phi$; $K$, $G_3$ and $G_4$ are free functions of $\phi$ and $X$.\footnote{For the sake of brevity we now set in what follows $G_i \equiv G_i\left(\phi,X\right)$, $G_{i,X} \equiv G_{iX} \equiv \frac{\partial G_i}{\partial X}$ and $G_{i,\phi} \equiv G_{i\phi} \equiv \frac{\partial G_i}{\partial \phi}$ where $i=2,3,4$.} Finally, we also assume $\mathcal{L}_\mathrm{m}$ includes the matter fields. 

These specific terms correspond to different dynamics, for example $K(\phi,X)$ contains the k-essence and quintessence theory, however does not contribute to the perturbations \cite{Frusciante:2018jzw}. On the other hand, the term $G_3(\phi,X)$ contains the so-called kinetic gravity braiding (KGB) models, with $G_{3X}\neq 0$ corresponding to mixing of the kinetic term of the scalar and the metric, with $G_3(\phi)$ only modifying the background as a dynamical DE. The last term, namely $G_4$, in in fact the one that contains the non-minimal coupling of the scalar to the Ricci curvature, and contains most scalar-tensor type theories. To give some concrete examples, the action \eqref{eq:action11} can easily be seen to reduce to the following subclasses:
\begin{itemize}
\item  \textbf{f(R) theories:} These are equivalent to a non-minimally coupled scalar field written as \cite{Chiba:2003ir}
\bea
K &=& -\frac{Rf_{,R}-f}{2\kappa}, \label{eq:fR-horndeski-1}\\
G_4 &=&\frac{\phi}{2\sqrt{\kappa}}, \label{eq:fR-horndeski}
\eea
where $\phi \equiv \frac{f_{,R}}{\sqrt{\kappa} }$ has units of mass and we have set $f_{,R} \equiv \dfrac{df}{dR}$.
\item \textbf{Brans-Dicke theories:} These are the archetype of a scalar-tensor theory, with
\bea
K &=& \frac{\omega_{BD} X}{\phi \sqrt{\kappa}}-V(\phi), \\
G_4 &=& \frac{\phi}{2 \sqrt{\kappa}},
\eea
where $V(\phi)$ is the potential and $\omega_{BD}$ is the well-known Brans-Dicke parameter \cite{Brans:1961sx}.
\item \textbf{Kinetic gravity braiding:} These models contain a  mixing of the scalar and tensor kinetic terms \cite{Deffayet:2010qz} and are given by
\bea
K &=&K(X), \\
G_3 &=& G_3(X), \\
G_4 &=& \frac{1}{2\kappa}.
\eea
\end{itemize}

\begin{itemize}
\item \textbf{The non-minimal coupling model:} This is given in terms of a coupling constant $\zeta$ as \cite{Quiros:2019ktw}
\bea
K &=& \omega(\phi)X-V(\phi),\\
G_4 &=& \left(\frac{1}{2\kappa}-\frac{\zeta \phi^2}{2}\right), \\
G_3 &=& 0. 
\eea
In the case of inflation, the Higgs-like inflation model is given by $\omega(\phi)=1$ and $V(\phi)=\lambda\left(\phi^2-\nu^2\right)^2/4$.
\item \textbf{Cubic Galileon:} The well-known models are given by  \cite{Quiros:2019ktw}
\bea
K &=& -X,\\
G_3 &\propto&  X,\\ 
G_4 &=& \frac{1}{2\kappa},
\eea
\end{itemize}

Then, varying the action of Eq.~\eqref{eq:action11} with respect to the metric and the scalar field $\phi$ we can obtain the equations of motion. First, doing a variation we find \cite{Kobayashi:2011nu}:
\bea
& & \delta \left(\sqrt{-g}\sum^{4}_{i=2} \mathcal{L}_i\right)=\sqrt{-g}\left[\sum^{4}_{i=2}\mathcal{G}^{i}_{\mu \nu}\delta g^{\mu \nu} +\sum^{4}_{i=2}\left(P_{\phi}^i-\nabla^{\mu}J_{\mu}^i\right)\delta \phi\right]+ \text{total deriv.},
\eea
from which the field equations follow. The gravitational field equations are:
\be
\label{eq:aa}
\sum^{4}_{i=2}\mathcal{G}^{i}_{\mu \nu}=\frac{1}{2} T_{\mu \nu}^{(m)},
\ee
where we have set
\bea
\mathcal{G}^2_{\mu \nu}&\equiv&-\frac{1}{2}K_{X}\nabla_{\mu}\phi\nabla_{\nu}\phi-\frac{1}{2}Kg_{\mu \nu}, \label{eq:def2}\\
\mathcal{G}^3_{\mu \nu}&\equiv&\frac{1}{2}G_{3X}\Box\phi\nabla_{\mu}\phi\nabla_{\nu}\phi+\nabla_{(\mu}G_3\nabla_{\nu)}\phi -\frac{1}{2}g_{\mu \nu}\nabla_{\lambda}G_3\nabla^{\lambda}\phi,\\
\mathcal{G}^4_{\mu \nu}&\equiv&G_4G_{\mu \nu}+g_{\mu \nu}\left(G_{4\phi}\Box\phi-2XG_{4\phi\phi}\right)-G_{4\phi}\nabla_{\mu}\nabla_{\nu}\phi  - G_{4\phi\phi}\nabla_{\mu}\phi\nabla_{\nu}\phi, \label{eq:def4}
\eea
and $T_{\mu \nu}^{(m)}$ is the energy-momentum tensor of matter. When $K=G_3=0$ and $G_4=\frac{1}{2\kappa}$ we can see that the Eqs.~\eqref{eq:aa} reduce to those of GR. Similarly, the equations of motion of the scalar field are given by
\begin{equation}
\label{eq:abb}
\nabla^{\mu}\left(\sum^4_{i=2}J^i_{\mu}\right)=\sum^{4}_{i=2}P^i_{\phi},
\end{equation}
where again we have set 
\bea
P^2_{\phi}&\equiv&K_{\phi}\\
P^3_{\phi}&\equiv&\nabla_{\mu}G_{3\phi}\nabla^{\mu}\phi,\\
P^4_{\phi}&\equiv&G_{4\phi}R, \\
J^2_{\mu}&\equiv&-\mathcal{L}_{2X}\nabla_{\mu}\phi\\
J^3_{\mu}&\equiv&-\mathcal{L}_{3X}\nabla_{\mu}\phi+G_{3X}\nabla_{\mu}X+2G_{3\phi}\nabla_{\mu}\phi,\\
J^4_{\mu}&\equiv&0.
\eea
Here it should be noted that while it would seem that the term $\nabla^{\mu}J^i_{\mu}$ leads to higher than  second-order derivatives, in fact it does not as was first noted in Ref.~\cite{Kobayashi:2011nu}. In fact this is due to that the commutations of higher derivatives can be shown to cancel out as 
\bea
\label{eq:comm}
& & \nabla_{\mu}\left(\Box \phi \nabla^{\mu}\phi+\nabla^{\mu}X\right)=\left(\Box \phi\right)^2-\left(\nabla_{\alpha}\nabla_{\beta}\phi\right)^2 -R_{\mu \nu}\nabla^{\mu}\phi \nabla^{\nu}\phi.
\eea
Doing some algebra it is possible to show that the scalar field equation \eqref{eq:abb} can be reduced to \cite{Arjona:2018jhh}
\bea
& & -\nabla_{\mu}K_{X}\nabla^{\mu}\phi - K_{X}\Box \phi - K_{\phi} + 2G_{3\phi}\Box \phi + \nabla_{\mu}G_{3\phi}\nabla^{\mu}\phi \nn \\
& & +\nabla_{\mu}G_{3X}\Box \phi \nabla^{\mu}\phi + \nabla_{\mu}G_{3X}\nabla^{\mu}X + G_{3X}\left[\left(\Box \phi\right)^2 - \right. \nn \\
& & \left. \left(\nabla_{\alpha}\nabla_{\beta}\phi\right)^2 - R_{\mu \nu}\nabla^{\mu}\phi \nabla^{\nu}\phi\right]-G_{4\phi}R=0.
\label{eq:scalar-field-equation-horndeski}
\eea
For the sake or brevity, in what follows we will denote by $X$ the kinetic term of the scalar field evaluated at the background and by $\delta X$ its linear order perturbation.

\subsubsection{Background expansion}
At the background level, assuming an unperturbed flat FRLW metric, it is easy to show that the modified Friedmann equations are given by
\bea
\mathcal{E}\equiv \sum^{4}_{i=2}\mathcal{E}_i&=&-\rho_\mathrm{m},\label{eq:epsi}\\
\mathcal{P}\equiv \sum^{4}_{i=2}\mathcal{P}_i&=&0,\label{eq:pp}
\eea
where we have defined the quantities
\begin{align}
\mathcal{E}_2 &\equiv 2XK_{X}-K, \\
\mathcal{E}_3 &\equiv 6X\dot{\phi}HG_{3X}-2XG_{3\phi}, \\
\mathcal{E}_4 &\equiv -6H^2G_4-6H\dot{\phi}G_{4\phi}, \\
\mathcal{P}_2 &\equiv K, \\
\mathcal{P}_3 &\equiv -2X\left(G_{3\phi}+\ddot{\phi}G_{3X}\right), \\
\mathcal{P}_4 &\equiv 2\left(3H^2+2\dot{H}\right)G_4+2\left(\ddot{\phi}+2H\dot{\phi}\right)G_{4\phi}+2\dot{\phi}^2G_{4\phi\phi}.
\end{align}
Gathering all the terms we can write down the explicit equations as 
\bea
& & 2XK_{X}-K+6X\dot{\phi}HG_{3X}-2XG_{3\phi}-6H^2G_4 -6H\dot{\phi}G_{4\phi}+\rho_\mathrm{m}=0,
\label{eq:friedmann1-horndeski}\\
& & K - 2X\left(G_{3\phi} + \ddot{\phi}G_{3X}\right) + 2\left(3H^2+2\dot{H}\right)G_4 +2\left(\ddot{\phi}+2H\dot{\phi}\right)G_{4\phi}+2\dot{\phi}^2G_{4\phi\phi}=0.
\label{eq:friedmann2-horndeski}
\eea
Again, in the limit of  $K=G_3=0$ and $G_4=\frac{1}{2\kappa}$, these reduced to the standard Friedmann equations as expected. By rearranging and collecting the terms in Eqs. \eqref{eq:friedmann1-horndeski}-\eqref{eq:friedmann2-horndeski} we can define the density parameter of an effective DE fluid
\bea
\label{eq:dde}
& & \bar{\rho}_\mathrm{DE} = \dot{\phi}^2K_{X}- K + 3\dot{\phi}^3HG_{3X} -\dot{\phi}^2G_{3\phi}  + 3H^2\left(\frac{1}{\kappa}-2G_{4}\right)-6H\dot{\phi}G_{4\phi},
\eea
and its effective pressure:
\bea
\label{eq:ppe}
& & \bar{P}_\mathrm{DE}= K - \dot{\phi}^2\left(G_{3\phi} + \ddot{\phi}G_{3X}\right)+ 2\dot{\phi}^2G_{4\phi\phi}  + 2\left(\ddot{\phi} +  2H\dot{\phi}\right)G_{4\phi} - \left(3H^2 + 2\dot{H}\right)\left(\frac{1}{\kappa}-2G_{4}\right),~~~~~~~~~
\eea
thus reducing the modified Friedmann equations Eqs. \eqref{eq:friedmann1-horndeski}-\eqref{eq:friedmann2-horndeski} to their traditional GR form
\begin{align}
3H^2&=\kappa\left(\bar{\rho}_\mathrm{DE}+\rho_\mathrm{m}\right), \label{eq:327} \\
 -\left(2\dot{H}+3H^2\right)&=\kappa\bar{P}_\mathrm{DE} ,
\end{align}
Using  Eqs. \eqref{eq:dde}-\eqref{eq:ppe} then allows us to also define the effective DE equation of state as
\begin{equation}
\label{eq:ww}
    w_\mathrm{DE}=\frac{K-\dot{\phi}^2\left(G_{3\phi}+\ddot{\phi}G_{3X}\right)-\left(3H^2+2\dot{H}\right)\left(\frac{1}{\kappa}-2G_{4}\right)+2\left(\ddot{\phi}+2H\dot{\phi}\right)G_{4\phi}+2\dot{\phi}^2G_{4\phi\phi}}{\dot{\phi}^2K_{X}-K+3\dot{\phi}^3HG_{3X}-\dot{\phi}^2G_{3\phi}+3H^2\left(\frac{1}{\kappa}-2G_{4}\right)-6H\dot{\phi}G_{4\phi}}.
\end{equation}

We can also write down the explicit scalar field equation \eqref{eq:scalar-field-equation-horndeski} as \cite{Kimura:2010di}
\bea
\label{eq:scalar-field-equation-zeroorder}
& & K_{\phi}-\left(K_{X}-2G_{3\phi}\right)\left(\ddot{\phi}+3H\dot{\phi}\right)-K_{\phi X}\dot{\phi}^2 - K_{XX}\ddot{\phi}\dot{\phi}^2 + G_{3\phi\phi}\dot{\phi}^2 + G_{3\phi X}\dot{\phi}^2\left(\ddot{\phi} - 3H\dot{\phi}\right) - \nn \\
& & 3G_{3X}\left(2H\dot{\phi}\ddot{\phi} + 3H^2\dot{\phi}^2+\dot{H}\dot{\phi}^2\right) - 3G_{3XX}H\dot{\phi}^3\ddot{\phi}  + 6G_{4\phi}\left(2H^2+\dot{H}\right)=0,
\eea
and by setting 
\bea
J_{\mu} &\equiv & \sum_{i=2}^4 J^i_{\mu}, \\
P_{\phi} & \equiv & \sum^{4}_{i=2}P^i_{\phi}
\eea
it is possible to rewrite the scalar field equation \eqref{eq:abb} as a conservation equation
\begin{equation}
\nabla_{\mu}J^{\mu}=P_{\phi},
\end{equation}
from which it is easy to deduce the existence of a Noether symmetry under constant shifts of the field $\phi \rightarrow \phi + c$, given by  \cite{Deffayet:2010qz}
\begin{equation}
J_{\mu}=\left(\mathcal{L}_{2X}+\mathcal{L}_{3X}-2G_{3\phi}\right)\nabla_{\mu}\phi-G_{3X}\nabla_{\mu}X.
\end{equation}
using the fact that that $X=\frac{1}{2}\dot{\phi}^2$, then we find that the conserved quantity is given by 
\begin{equation}
\label{eq:jj}
    J \equiv J_0=\dot{\phi}\left(K_X-2G_{3\phi}+3H\dot{\phi}G_{3X}\right),
\end{equation}
so that the scalar field equation reduces to a much simpler form
\begin{equation}
    \dot{J}+3HJ=P_{\phi}.
\end{equation}
When $P_{\phi}=0$, then the solution is simply
\be
J=\frac{J_c}{a^3},
\ee
for a constant $J_c$. Here we can also classify some particular subcases: first, when $J_c=0$ then the scalar field is on the attractor solution, while when $J_c\neq0$ then the system is not on the attractor and new dynamics may arise \cite{Arjona:2018jhh}.

\subsubsection{Linear perturbations}
By using the perturbed FLRW metric of Eq.~\eqref{eq:FRWpert} with the  field equations \eqref{eq:aa} we can obtain the linear theory predictions for the perturbations  \cite{DeFelice:2011hq,Matsumoto:2018dim}
\bea
& & A_1\dot{\Phi}+A_2\dot{\delta \phi}+A_3\frac{k^2}{a^2}\Phi+A_4\Psi +\left(A_6\frac{k^2}{a^2} -\mu \right)\delta \phi -\rho_\mathrm{m}\delta_\mathrm{m}=0, \label{eq:1}\\
& & C_1\dot{\Phi}+C_2\dot{\delta \phi}+C_3\Psi+C_4\delta \phi-\frac{a \rho_\mathrm{m} V_\mathrm{m}}{k^2}=0, \label{eq:field-equation-horndeski-3}\\
& & B_1\ddot{\Phi}+B_2\ddot{\delta \phi}+B_3\dot{\Phi}+B_4\dot{\delta \phi}+B_5\dot{\Psi}+B_6\frac{k^2}{a^2}\Phi  +\left(B_7\frac{k^2}{a^2}+3\nu\right)\delta \phi+\left(B_8\frac{k^2}{a^2}+B_9\right)\Psi=0, \label{eq:2}~~~~~~~~~\\
& & G_4\left(\Psi+\Phi\right)+G_{4\phi}\delta \phi=0. \label{eq:field-equation-horndeski-4}
\eea
Again, in the case when $K = G_3 = 0$ and $G_4 = \frac{1}{2\kappa}$, we can see that  Eqs.~\eqref{eq:1}-\eqref{eq:field-equation-horndeski-4} reduce to the GR limit given by Eqs.~\eqref{eq:phiprimeeq}-\eqref{eq:anisoeq} with no anisotropic stress. On the other hand, if we consider  Eq.~\eqref{eq:scalar-field-equation-horndeski} we similarly find the perturbed equations of motion for the scalar field
\bea
\label{eq:3}
& & D_1\ddot{\Phi}+D_2\ddot{\delta \phi}+D_3\dot{\Phi}+D_4\dot{\delta \phi}+D_5\dot{\Psi}+\left(D_7\frac{k^2}{a^2}+D_8\right)\Phi +\left(D_9\frac{k^2}{a^2}-M^2\right)\delta \phi \nn \\
& & +\left(D_{10}\frac{k^2}{a^2}+D_{11}\right)\Psi=0,
\eea
where the full expressions for the variables $A_i$, $\mu$, $M$, $\nu$, $B_i$, $C_i$ and $D_i$ can be found in Appendix B of Ref.~\cite{Arjona:2019rfn}.

\subsubsection{The effective fluid approach for Horndeski models \label{Section:EFA}}

Following a similar approach as in the case of the $f(R)$ models, we can now apply the effective fluid approach to the Horndeski models as well. While we already showed this for the background  effective DE density and pressure given by Eqs.~\eqref{eq:dde}-\eqref{eq:ppe}, so in what follows we also do the same for the linear theory of these models, under the sub-horizon and quasi-static approximations. 

Obviously, the first step is to define an effective DE fluid by moving all MG contributions to the right hand side of the field equations and define the DE effective energy-momentum tensor $T_{\mu\nu}^\mathrm{DE}$. Doing so with the gravitational field equations \eqref{eq:aa}, we find:
\bea
G_{\mu\nu}&=&\kappa\left(T_{\mu\nu}^{(m)}+T_{\mu\nu}^{DE}\right),\nn \\
\kappa T_{\mu\nu}^\mathrm{DE}&=&G_{\mu\nu}-2\kappa \sum^{4}_{i=2}\mathcal{G}^{i}_{\mu \nu}.
\eea
By considering the decomposition of the  $T_{\mu\nu}^\mathrm{DE}$ tensor into its components, given by  Eqs.~\eqref{eq:effectTmn1}-\eqref{eq:effectTmn}, we can extract the expressions for the DE effective perturbations in the pressure, density, and velocity. Doing so, we find the latter have the general structure:
\bea
\frac{\delta P_\mathrm{DE}}{\bar{\rho}_\mathrm{DE}}&=&(\ldots)\delta \phi+(\ldots)\dot{\delta \phi}+(\ldots)\ddot{\delta \phi}+(\ldots)\Psi +(\ldots)\dot{\Psi} \nn \\
&+&(\ldots)\Phi +(\ldots)\dot{\Phi}+(\ldots)\ddot{\Phi},\label{eq:effdenp0}\\
\delta_\mathrm{DE}&=&(\ldots)\delta \phi+(\ldots)\dot{\delta \phi}+(\ldots)\Psi + (\ldots)\Phi
+(\ldots)\dot{\Phi},\label{eq:effprp01} \\
V_\mathrm{DE}&=&(\ldots)\delta \phi+(\ldots)\dot{\delta \phi}+(\ldots)\Psi +(\ldots)\Phi
+(\ldots)\dot{\Phi}.\label{eq:efftheta01}
\eea
where  the dots $(\ldots)$ indicates long expressions.

Following the same procedure as before and using the sub-horizon and quasi-static approximations to Eqs.~\eqref{eq:1}, \eqref{eq:2}, and Eq.~\eqref{eq:3} we find \cite{Arjona:2019rfn}
\bea
\label{eq:bla1}
& & A_3\frac{k^2}{a^2}\Phi+A_6\frac{k^2}{a^2}\delta \phi-\kappa \rho_\mathrm{m}\delta_\mathrm{m} \simeq 0,\\
\label{eq:trace}
& & B_6\frac{k^2}{a^2}\Phi +  B_8\frac{k^2}{a^2}\Psi + B_7\frac{k^2}{a^2}\delta \phi  \simeq 0,\\
\label{eq:bla}
& & D_7\frac{k^2}{a^2}\Phi+\left(D_9\frac{k^2}{a^2}-M^2\right)\delta \phi+D_{10}\frac{k^2}{a^2}\Psi \simeq 0.
\eea
As $B_7=4G_{4\phi}$ and $B_6=B_8$ (see for example Appendix B of Ref.~\cite{Arjona:2019rfn}), then we see that Eq.~(\ref{eq:trace}) leads to $\Phi=-\Psi$ when $G_4$ is a constant, implying no DE anisotropic stress. Then by solving  Eqs.~\eqref{eq:bla1}-\eqref{eq:bla} for $\Phi$, $\Psi$ and $\delta \phi$ we obtain the Poisson equations 
\bea
\frac{k^2}{a^2}\Psi&=&-\frac{\kappa}{2}\frac{G_{\textrm{eff}}}{G_\mathrm{N}}\bar{\rho}_\mathrm{m}\delta,\\
\frac{k^2}{a^2}\Phi&=&\frac{\kappa}{2} Q_{\textrm{eff}}\bar{\rho}_\mathrm{m}\delta, \\
\delta \phi&=&\frac{\left(A_6B_6-B_6B_7\right)\rho_\mathrm{m}\delta_\mathrm{m}}{\left(A^2_6B_6-2A_6B_6B_7+B^2_6D_9\right)\frac{k^2}{a^2}-B^2_6M^2},~~~
\eea
where the parameters $G_{\textrm{eff}}$ and $Q_{\textrm{eff}}$ are Newton's effective constant and the lensing variable
\bea
& & \frac{G_{\textrm{eff}}}{G_\mathrm{N}} =\frac{2\left[\left(B_6D_9-B^2_7\right)\frac{k^2}{a^2}-B_6M^2\right]}{\left(A^2_6B_6+B^2_6D_9-2A_6B_7B_6\right)\frac{k^2}{a^2}-B^2_6M^2},~~~~~~\label{eq:ODEGeff}\\
& & Q_{\textrm{eff}} =\frac{2\left[\left(A_6B_7-B_6D_9\right)\frac{k^2}{a^2}+B_6M^2\right]}{\left(A^2_6B_6+B^2_6D_9-2A_6B_7B_6\right)\frac{k^2}{a^2}-B^2_6M^2}.~~~~~~
\eea
We also define the DE anisotropic stress parameters as 
\begin{align}
    \eta &\equiv \frac{\Psi+\Phi}{\Phi}=\frac{\left(A_6-B_7\right)B_7\frac{k^2}{a^2}}{\left(A_6B_7-B_6D_9\right)\frac{k^2}{a^2}+B_6M^2},\\
    \gamma &\equiv -\frac{\Phi}{\Psi}=\frac{\left(A_6B_7-B_6D_9\right)\frac{k^2}{a^2}+B_6M^2}{\left(B^2_7-B_6D_9\right)\frac{k^2}{a^2}+B_6M^2},
\end{align}
which are in agreement with the ones found in Ref.~\cite{DeFelice:2011hq}. For these models Eq.~\eqref{eq:Geffode} is still valid, albeit with $G_\mathrm{eff}$ given by Eq.~\eqref{eq:ODEGeff}.

\subsection{Scalar-Vector-Tensor models 
\label{sec:svt}}
An interesting extension of Horndeski models is Scalar-Vector-Tensor (SVT) models, which also include a vector degree of freedom and also include generalised Proca theories\cite{Heisenberg:2018mxx}. The most general SVT Lagrangian is given by \cite{Cardona:2022lcz}
\begin{equation}
\mc{L} \equiv \sum_{i = 2}^6 \mc{L}_i^\text{SVT} + \sum_{i = 2}^5 \mc{L}_i^\text{ST} + \mc{L}_\mathrm{m}, \label{eq:L}
\end{equation}
where the Lagrangians $\mc{L}_i^\text{ST}$ include the scalar-tensor terms, while the Lagrangians $\mc{L}_i^\text{SVT}$ contain the scalar-vector-tensor terms. 

The SVT models in fact contain a scalar field $\vp$, a vector field $A_\mu$ and the gravitational field $g_{\mu\nu}$, which are related to each other via the $\mc{L}_i^\text{SVT}$ terms, given by
\begin{align}
\mc{L}_2^\text{SVT} &= f_2 (\vp, X_1, X_2, X_3, F, Y_1, Y_2, Y_3), \label{eq:svt2}\\
\mc{L}_3^\text{SVT} &= f_3 (\vp, X_3) g^{\mu\nu} S_{\mu\nu} + \tilde{f}_3 (\phi, X_3) A^\mu A^\nu S_{\mu\nu}, \\
\mc{L}_4^\text{SVT} &= f_4 (\vp, X_3) R + f_{4 X_3} (\vp, X_3) \left\{ \( \nabla_\mu A^\mu \)^2 - \nabla_\mu A_\nu \nabla^\nu A^\mu \right\}, \\
\mc{L}_5^\text{SVT} &= f_5 (\vp, X_3) G^{\mu\nu} \nabla_\mu A_\nu + \mc{M}_5^{\mu\nu} \nabla_\mu \nabla_\nu \vp +\mc{N}^{\mu\nu}_5 S_{\mu\nu} \nonumber \\
 &- \frac{1}{6} f_{5 X_3} (\vp, X_3) \left\{ \( \nabla_\mu A^\mu \)^3 - 3 \( \nabla_\mu A^\mu \) \nabla_\rho A_\sigma \nabla^\sigma A^\rho + 2 \nabla_\rho A_\sigma \nabla^\tau A^\rho \nabla^\sigma A_\tau \right\}, \\
\mc{L}_6^\text{SVT} &= f_6 (\vp, X_1) L^{\mu\nu\alpha\beta} F_{\mu\nu} F_{\alpha\beta} + \tilde{f}_6 (\vp, X_3) L^{\mu\nu\alpha\beta} F_{\mu\nu} F_{\alpha\beta}  \nonumber \\
 &+ \mc{M}^{\mu\nu\alpha\beta}_6 \nabla_\mu \nabla_\alpha \vp \nabla_\nu \nabla_\beta \vp + \mc{N}^{\mu\nu\alpha\beta}_6 S_{\mu\alpha} S_{\nu\beta}, \label{eq:svt6}
\end{align}
where for the sake of brevity we defined $g_{\xi} \equiv \tfrac{\partial g}{\partial \xi}$, denoting the derivative of a  $g$ with respect to the scalar $\xi$. In the previous equations, we also defined the kinetic terms and couplings between the scalar and vector fields as 
\begin{equation}
X_1 \equiv - \frac{1}{2} \nabla_\mu \vp \nabla^\mu \vp, \quad X_2 \equiv - \frac{1}{2} A_\mu \nabla^\mu \vp, \quad X_3 \equiv - \frac{1}{2} A_\mu A^\mu. \label{eq:x1x2x3}
\end{equation}
As usual, the antisymmetric tensor $F_{\mu\nu}$ and its dual $\tilde{F}^{\mu\nu}$ are related to $A_\mu$ via
\begin{equation}
F_{\mu\nu} \equiv \nabla_\mu A_\nu - \nabla_\nu A_\mu, \quad \tilde{F}^{\mu\nu} \equiv \frac{1}{2} \varepsilon^{\mu\nu\alpha\beta} F_{\alpha\beta},
\end{equation}
where $\varepsilon^{\mu\nu\alpha\beta} \equiv \frac{\epsilon^{\mu\nu\alpha\beta}}{\sqrt{-g}}$, $\epsilon^{\mu\nu\alpha\beta}$ is the Levi-Civita symbol. Furthermore, the Lorentz invariant quantities can be constructed from $F_{\mu\nu}$ as
\begin{align}
F &\equiv - \frac{1}{4} F_{\mu\nu} F^{\mu\nu}, \nn \\
Y_1 \equiv \nabla_\mu \vp \nabla_\nu \vp F^{\mu\alpha} F^\nu_{\ \alpha}, \qquad Y_2 &\equiv \nabla_\mu \vp A_\nu F^{\mu\alpha} F^\nu_{\ \alpha}, \qquad Y_3 \equiv A_\mu A_\nu F^{\mu\alpha} F^\nu_{\ \alpha}, \label{eq:lorentz-inv} 
\end{align} 
which vanish when $A_\mu \rightarrow \nabla_\mu \pi$, where $\pi$ is a scalar field. Moreover, the symmetric tensor $S_{\mu\nu}$ is related to $A_\mu$ as
\begin{equation}
S_{\mu\nu} \equiv \nabla_\mu A_\nu + \nabla_\nu A_\mu,
\end{equation}
while the tensors $\mc{M}$ and $\mc{N}$ are given by
\begin{equation}
\mc{M}^{\mu\nu}_5 \equiv \mc{G}_{\rho\sigma}^{h_5} \tilde{F}^{\mu\rho} \tilde{F}^{\nu\sigma}, \quad \mc{N}^{\mu\nu}_5 \equiv \mc{G}_{\rho\sigma}^{\tilde{h}_5} \tilde{F}^{\mu\rho} \tilde{F}^{\nu\sigma},
\end{equation}
\begin{equation}
\mc{M}^{\mu\nu\alpha\beta}_6 \equiv 2 f_{6 X_1} (\vp, X_1) \tilde{F}^{\mu\nu} \tilde{F}^{\alpha\beta}, \quad \mc{N}^{\mu\nu\alpha\beta}_6 \equiv \frac{1}{2} \tilde{f}_{6 X_3} (\vp, X_1) \tilde{F}^{\mu\nu} \tilde{F}^{\alpha\beta},
\end{equation}
where
\begin{align}
\mc{G}^{h_5}_{\rho\sigma} &\equiv h_{51} (\vp, X_i) g_{\rho\sigma} + h_{52} (\vp, X_i) \nabla_\rho \vp \nabla_\sigma \vp + h_{53} (\vp, X_i) A_\rho A_\sigma + h_{54} (\vp, X_i) A_\rho \nabla_\sigma \vp, \label{eq:Gh5}\\
\mc{G}^{\tilde{h}_5}_{\rho\sigma} &\equiv \tilde{h}_{51} (\vp, X_i) g_{\rho\sigma} + \tilde{h}_{52} (\vp, X_i) \nabla_\rho \vp \nabla_\sigma \vp + \tilde{h}_{53} (\vp, X_i) A_\rho A_\sigma + \tilde{h}_{54} (\vp, X_i) A_\rho \nabla_\sigma \vp. \label{eq:Ght5}
\end{align}
Finally, the $L^{\mu\nu\alpha\beta}$ tensor is given by
\begin{equation}
L^{\mu\nu\alpha\beta} \equiv \frac{1}{4} \varepsilon^{\mu\nu\rho\sigma} \varepsilon^{\alpha\beta\gamma\delta} R_{\rho\sigma\gamma\delta}. \label{eq:Lmunualphabeta}
\end{equation}

As always, the scalar-tensor interactions are contained in the Horndeski theory discussed in the previous section:
\begin{align}
\mc{L}_2^\text{ST} &= G_2 (\vp, X_1), \label{eq:st2}\\
\mc{L}_3^\text{ST} &= - G_3 (\vp, X_1) \square \vp, \\
\mc{L}_4^\text{ST} &= G_4 (\vp, X_1) R + G_{4 X_1} (\vp, X_1) \left\{ \( \square \vp \)^2 - \nabla_\mu \nabla_\nu \vp \nabla^\nu \nabla^\mu \vp \right\}, \\
\mc{L}_5^\text{ST} &= G_5 (\vp, X_1) G^{\mu\nu} \nabla_\mu \nabla_\nu \vp \label{eq:st5}\\
 &- \frac{1}{6} G_{5 X_1} (\vp, X_1) \left\{ \( \square \vp \)^3 - 3 \( \square \vp \) \nabla_\mu \nabla_\nu \vp \nabla^\nu \nabla^\mu \vp + 2 \nabla^\mu \nabla_\sigma \vp \nabla^\sigma \nabla_\rho \vp \nabla^\rho \nabla_\mu \vp \right\} \nonumber, 
\end{align}
where the terms $f_i$, $\tilde{f}_i$, $h_{5i}$, $\tilde{h}_{5i}$, and $G_i$ are free functions.  

It should be noted that even though this theory contains several free and a priori undetermined functions, in practice it has been significantly constrained by the recent GW discovery \cite{Abbott:2017oio}.

\subsubsection{The effective fluid approach for SVT theories with non-vanishing anisotropic stress 
\label{sec:without-piDE}}
Following the same methodology as before, the quasi-static and sub-horizon approximations can be applied to the SVT model in order to determine the DE fluid parameters in the effective fluid approach. These where found by Ref.~\cite{Cardona:2022lcz} to be
\begin{equation*}
\delta \rho_\text{DE} = \frac{\frac{k^6}{a^6} Z_1 + \frac{k^4}{a^4} Z_2 + \frac{k^2}{a^2} Z_3 + Z_4}{\frac{k^6}{a^6} Z_5 + \frac{k^4}{a^4} Z_6 + \frac{k^2}{a^2} Z_7} \delta \rho_\mathrm{m}, \quad \delta P_\text{DE} = \frac{1}{3 Z_{12}} \frac{\frac{k^6}{a^6} Z_8 + \frac{k^4}{a^4} Z_9 + \frac{k^2}{a^2} Z_{10} + Z_{11}}{\frac{k^6}{a^6} Z_5 + \frac{k^4}{a^4} Z_6 + \frac{k^2}{a^2} Z_7} \delta \rho_\mathrm{m},
\end{equation*}
\begin{equation*}
\frac{a \bar{\rho}_\text{DE}}{k^2} V_\text{DE} = \frac{\frac{k^4}{a^4} Z_{13} + \frac{k^2}{a^2} Z_{14} + Z_{15}}{\frac{k^6}{a^6} Z_5 + \frac{k^4}{a^4} Z_6 + \frac{k^2}{a^2} Z_7} \delta \rho_\mathrm{m}, \quad \bar{\rho}_\text{DE} \pi_\text{DE} = \frac{k^2}{a^2} \frac{\tfrac{k^2}{a^2} (W_{14} - W_{11}) + ( W_{15} - W_{12})}{\tfrac{k^4}{a^4} W_3 + \tfrac{k^2}{a^2} W_4 + W_5} \delta \rho_\mathrm{m},
\end{equation*}
\begin{equation}
c_\mathrm{s, \text{DE}}^2 = \frac{1}{3 Z_{12}} \frac{\tfrac{k^6}{a^6} Z_8 + \tfrac{k^4}{a^4} Z_9 + \tfrac{k^2}{a^2} Z_{10} + Z_{11}}{\tfrac{k^6}{a^6} Z_1 + \tfrac{k^4}{a^4} Z_2 + \tfrac{k^2}{a^2} Z_3 + Z_4},
\label{Stress Eqs}
\end{equation}
where the coefficients $Z_i$ ($i = 1,\dots,15$) are given in  Appendix E of \cite{Cardona:2022lcz}. As in previous cases, the sound speed given in Eq.~\eqref{Stress Eqs} does not on its own determine the stability of sub-horizon perturbations, but in fact the relevant quantity is the effective sound speed given, as in the previous cases, by the difference of the DE sound speed and a term proportional to the anisotropic stress $\pi = \frac{3}{2} (1 + w) \sigma$. The effective sound speed again in this case is defined as \cite{Cardona:2014iba}
\be 
c_\mathrm{s, \text{eff}}^2 \equiv c_\mathrm{s, \text{DE}}^2 - \frac{2}{3} \frac{\bar{\rho}_\text{DE} \pi_\text{DE}}{\delta \rho_\text{DE}}.
\ee 

\section{The effective fluid approach and the Boltzmann codes \label{sec:apps}}
\subsection{Designer Horndeski}
One particularly interesting model to demonstrate the effective fluid approach is a class of designer Horndeski parameterization, discovered in Ref.~\cite{Arjona:2019rfn} and rediscovered later in Ref.~\cite{Linder:2021est}. These models have a background expansion exactly equal to that of the \lcdm model, but also have different perturbations, and are particularly useful in searches for deviations from \lcdm \cite{Nesseris:2013fca,Arjona:2018jhh}. 

For example, in a Horndeski model with just the $G_2$ and $G_3$ terms, i.e. of the KGB type, we can use the modified Friedmann equation 
\bea
\label{eq:friedeq}
& & -H(a)^2-\frac{K(X)}{3}+H^2_0\Omega_\mathrm{m}(a)+2\sqrt{2}X^{3/2}H(a)G_{3X}  +\frac{2}{3}X K_X=0.
\eea
and the scalar field conservation equation
\begin{equation}
\label{eq:scfeq}
    \frac{J_c}{a^3}-6XH(a)G_{3X}-\sqrt{2}\sqrt{X}K_X=0
\end{equation}
to solve for the unknown functions, while demanding that $H(a)$ corresponds to the of the \lcdm model. In the previous equations $J_c$ is a constant which quantifies our deviation from the attractor, as in the case of the KGB model \cite{Kimura:2010di}. Solving Eq.~\eqref{eq:friedeq} and \eqref{eq:scfeq} for $(G_{3X}(X),K(X))$ yields \cite{Arjona:2019rfn}:
\bea
\label{eq:systemdes}
K(X) &=& -3 H_0^2 \Omega_{\Lambda,0}+\frac{J_c \sqrt{2X} H(X)^2}{H_0^2 \Omega_\mathrm{m,0}}-\frac{J_c \sqrt{2X} \Omega_{\Lambda,0}}{\Omega_\mathrm{m,0}} \nn\\
G_{3X}(X) &=& -\frac{2 J_c H'(X)}{3 H_0^2 \Omega_\mathrm{m,0}},
\eea
where we have written $H=H(X)$, i.e. in terms of the kinetic term. in fact, Eqs.~\eqref{eq:systemdes} give us a whole family of designer models that behave as \lcdm at the background level but have different perturbations. For example, assuming that $X= \frac{c_0}{H(a)^n}$, where $c_0>0$ and $n>0$, we find \cite{Arjona:2019rfn}:
\bea
\label{eq:bestdes}
G_{3}(X)&=&-\frac{2 J_c c_0^{1/n} X^{-1/n}}{3 H_0^2 \Omega_\mathrm{m,0}},\\
K(X)&=&\frac{\sqrt{2} J_c c_0^{2/n} X^{\frac{1}{2}-\frac{2}{n}}}{H_0^2 \Omega_\mathrm{m,0}}-3 H_0^2 \Omega_{\Lambda,0}-\frac{\sqrt{2} J_c \sqrt{X} \Omega_{\Lambda,0}}{\Omega_\mathrm{m,0}}.\nn
\eea
As can be seen, this designer model, designated as \textit{HDES} hereafter, has a nice and smooth limit to the  \lcdm model and it also recovers GR when $J_c\sim0$.

\subsection{Numerical solutions of the perturbation equations}
Here we now present the numerical solutions of the perturbation equations in the effective fluid approach, using as an example the HDES model, given by Eq.~\eqref{eq:bestdes}. In particular:
\begin{itemize}
  \item First, we consider the numerical solution of the full system of equations given by Eqs.~\eqref{eq:1}-\eqref{eq:field-equation-horndeski-4}, which we call ``Full-DES".
  \item Second, we consider the numerical solution of the effective fluid approach given by Eqs.~\eqref{Eq:evolution-delta}-\eqref{Eq:evolution-V}, which we call  ``Eff. Fluid".
  \item Third, we consider the numerical solution of the growth factor equation \eqref{eq:Geffode} using the appropriate expression for $G_\mathrm{eff}$, which we call  ``ODE-Geff".
  \item Finally, we also consider the \lcdm model.
\end{itemize}

\begin{figure*}[!t]
\centering
\includegraphics[width=0.49\textwidth]{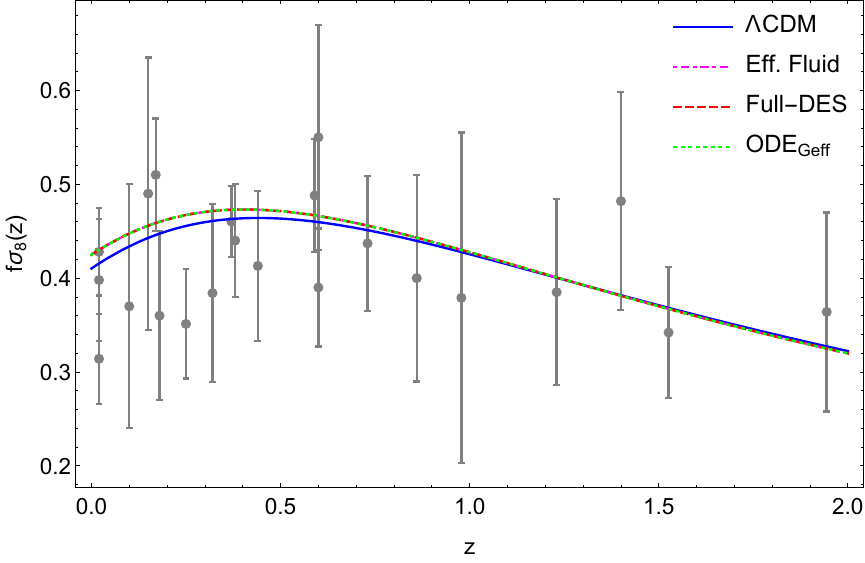}
\includegraphics[width=0.49\textwidth]{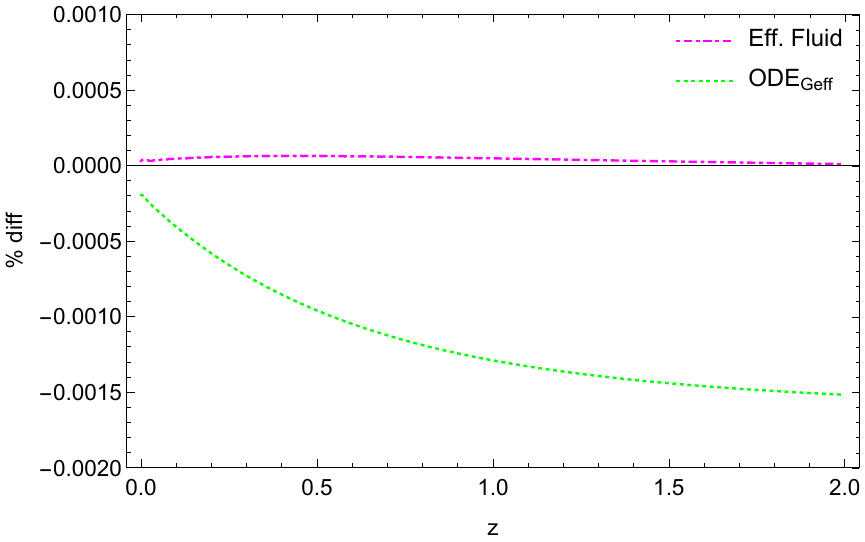}
\caption{Left: The theoretical predictions for the $f\sigma_8(z)$ parameter of the HDES model with $n=2$, $\tilde{J}_c=5\cdot10^{-2}$ and $\sigma_{8,0}=0.8$ versus the $f\sigma_8$ data compilation from Ref.~\cite{Sagredo:2018ahx}. Right: The percent difference between the ``Full-DES" brute-force numerical solution and the effective fluid approach (magenta dot dashed line) and the numerical solution of the growth factor equation \eqref{eq:Geffode} (green dotted line). Image from Ref.~\cite{Arjona:2019rfn}.\label{fig:HDESevo}}
\end{figure*}

As a concrete example, we set $\tilde{c}_0=1$, $n=2$, $\Omega_\mathrm{m,0}=0.3$, $k=300H_0$ and $\sigma_{8,0}=0.8$, unless otherwise specified. Then we show the evolution of the growth-rate parameter $f\sigma_8(z)$ for the HDES model on the left panel of Fig.~\ref{fig:HDESevo}. Specifically, we show the ``Full-DES" brute-force numerical  solution, the effective fluid approach, the $\Lambda$CDM model and the numerical solution of the $G_{\textrm{eff}}$ equation and as can be seen, the agreement between all approaches is excellent. On the other hand, in the right panel of  Fig.~\ref{fig:HDESevo} we show the percent difference between the ``Full-DES" brute-force numerical solution and the effective fluid approach (magenta dot dashed line) and the numerical solution of the growth factor equation \eqref{eq:Geffode} (green dotted line).

\subsection{Modifications to CLASS and the ISW effect.}
Finally, we now show how the effective fluid approach can be implemented into a Boltzmann code, like CLASS \cite{Blas:2011rf}. Our modifications to the code are denoted as EFCLASS \cite{Arjona:2018jhh,Arjona:2019rfn}, while we also compare with the hi\_CLASS code \cite{Zumalacarregui:2016pph}, which solves the full set of dynamical equations, but at the cost of significantly more complicated modifications. 

In our case however, the modifications for the effective fluid approach are much easier, as we only require the DE velocity and the anisotropic stress \cite{Arjona:2018jhh,Arjona:2019rfn}. In the particular case where we consider the HDES model, then there is the further simplification that the anisotropic stress $\pi_\mathrm{DE}$ is also zero, as can be seen from Eq.~(\ref{eq:field-equation-horndeski-4}), since $G_{4\phi}=0$. Thus, to modify CLASS we only require the expression for the DE velocity, which for $n=1$ is given by 
\be
V_\mathrm{DE}\simeq \left(-\frac{14 \sqrt{2}}{3} \Omega_\mathrm{m,0}^{-3/4} \tilde{J_c}~H_0~ a^{1/4}\right)\frac{\bar{\rho}_\mathrm{m}}{\bar{\rho}_\mathrm{DE}} \delta_\mathrm{m}.\label{eq:VDEHDES}
\ee

\begin{figure*}[!t]
\centering
\includegraphics[width=0.49\textwidth]{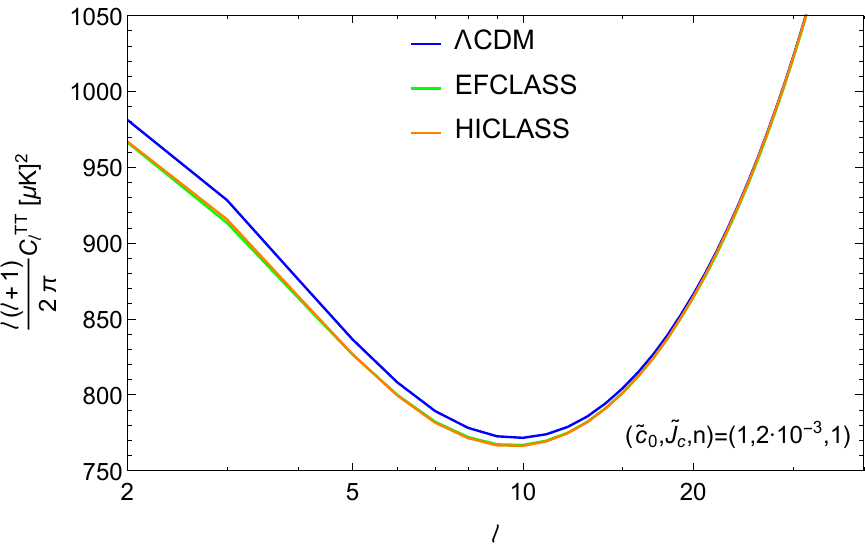}
\includegraphics[width=0.49\textwidth]{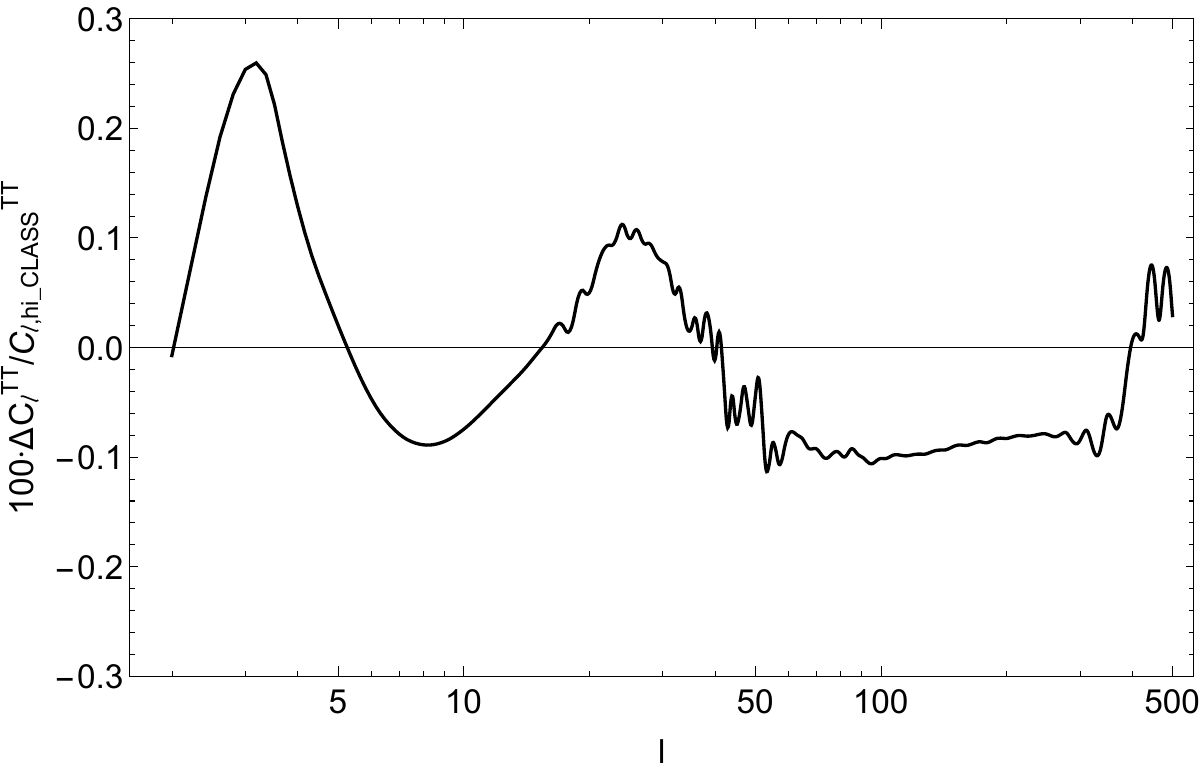}
\caption{Left: The TT CMB spectrum for $\Omega_\mathrm{m,0}=0.3$, $n_s=1$, $A_s=2.3 \cdot 10^{-9}$, $h=0.7$ and $(\tilde{c_0},\tilde{J_c},n)=(1,2\cdot 10^{-3},1)$. The results of the EFCLASS code are denoted by the green line, those of hi\_CLASS by an orange line and for the \lcdm with a blue line. Right: The percent difference of our code with hi\_CLASS. Image from Ref.~\cite{Arjona:2019rfn}. \label{fig:classcls}}
\end{figure*}

Next, we show the results of applying Eq.~\eqref{eq:VDEHDES} to the CLASS code and comparing with hi\_CLASS. This is shown in the  left panel of Fig.~\ref{fig:classcls}, where the low-$\ell$ multipoles of the TT CMB spectrum for a flat universe with $\Omega_\mathrm{m,0}=0.3$, $n_s=1$, $A_s=2.3 \cdot 10^{-9}$, $h=0.7$ and $(\tilde{c_0},\tilde{J_c},n)=(1,2\cdot 10^{-3},1)$ can be seen. Our EFCLASS code is denoted by the green line, hi\_CLASS is given by the orange line, while the blue line  corresponds to the \lcdm model. As can be seen, the agreement between EFCLASS and hi\_CLASS is remarkable and even with our simple modification is roughly $\sim 0.1\%$ accuracy across all multipoles (shown on the right panel of Fig.~\ref{fig:classcls}).

Finally, we also compare our modifications of the effective fluid approach with a direct calculation of the Integrated Sachs-Wolfe (ISW) effect. Specifically, the temperature power spectrum is given by \cite{Song:2006ej}:
\bea
C_\ell^{\textrm{ISW}}=4\pi \int \frac{dk}{k} I_\ell^{\textrm{ISW}}(k)^2 \frac{9}{25} \frac{k^3 P_{\zeta}(k)}{2\pi^2},\label{eq:clsISWtheory}
\eea
where $I_\ell^{\textrm{ISW}}(k)$ is a kernel that depends on the line of sight integral of the growth and a bessel function, while $P_{\zeta}(k)$ is the primordial power spectrum, given by the primordial power spectrum times a transfer function \cite{Song:2006ej,Arjona:2018jhh}
\be
\frac{k^3 P_{\zeta}}{2\pi^2}=A_s \left(\frac{k}{k_0}\right)^{n_s-1} T(k)^2,
\ee
where $A_s$ is the primordial amplitude, $k_0$ is the pivot scale and $T(k)$ is the Bardeen, Bond, Kaiser and Szalay (BBKS) transfer function \cite{Dodelson:2003ft}.

\begin{figure*}[!t]
\centering
\includegraphics[width=0.49\textwidth]{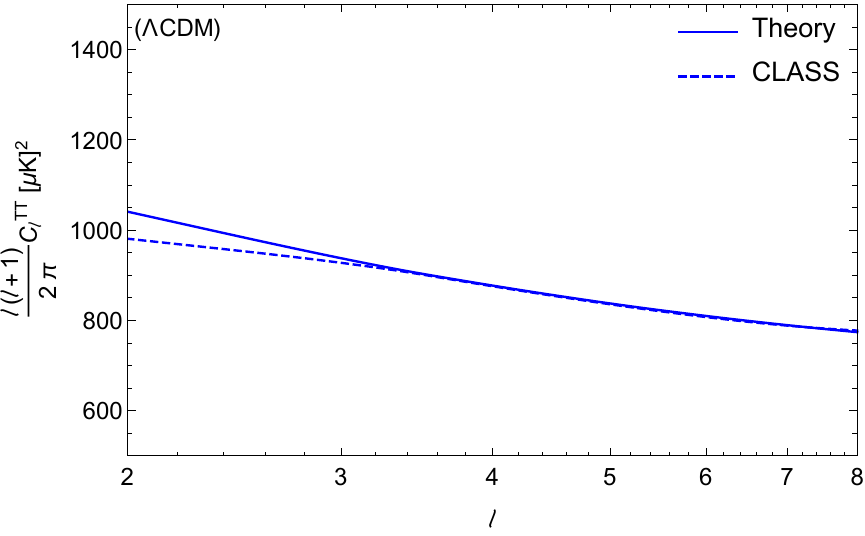}
\includegraphics[width=0.49\textwidth]{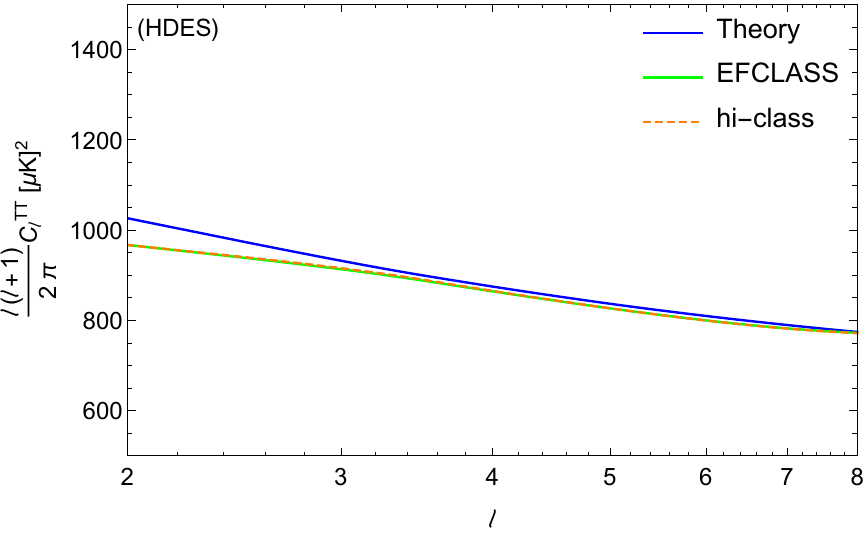}
\caption{Plots of the ISW effect and a comparison with CLASS/hi\_CLASS for the \lcdm model (left) and the HDES model (right), for the same parameters as in Fig.~\ref{fig:classcls}. Image from Ref.~\cite{Arjona:2019rfn}. \label{fig:classclsisw}}
\end{figure*}

We show in Fig.~\ref{fig:classclsisw} a comparison between  CLASS and hi\_CLASS for the \lcdm model (left) and the HDES models (right), for the same parameters as in Fig.~\ref{fig:classcls}. Overall, there is good agreement at all multipoles, except for $\ell=2$, as the BBKS formula is only $10\%$ accurate on large scales or equivalently, small multipoles.

\section{Conclusions \label{sec:conc}}

In this review we briefly presented the so-called effective fluid approach, which is a framework that allows for treating modified gravity models as GR with an ideal DE fluid, described by an equation of state, a sound speed, an anisotropic stress and DE pressure, density and velocity perturbations. While in general MG models have very complicated evolution equations, they are significantly simplified under the effective fluid approach with the addition of the quasi-static and sub-horizon approximations, as they can be written in terms simple DE fluid equations. However, it should noted that in general it is not possible to discriminate modified gravity models from GR plus DE exotic fluids, as one may always move the modifications of gravity to the right hand side of the field equations and rewrite them as the Einstein equations with extra matter/DE contributions. 

Thus, the main advantage of this approach, as highlighted in this work, is that it allows us to easily including most MG models in the Boltzmann codes, as the latter are typically fine-tuned and hard-coded for GR and a DE fluid. Here, we presented the main formalism and the results for the DE fluid quantities for several MG models, including the $f(R)$, Horndeski and Scalar-Vector-Tensor models, in all of which we presented the critical quantities that describe the evolution of the perturbations and the effective DE fluid equations.

We also described some specific applications, i.e. the numerical solutions of the  fluid equations for growth factor, the designer Horndeski model and how our approach can be implemented in the Boltzmann codes. In the case of the HDES model we found that with our effective fluid approach and just a simple modification, the agreement between our modified EFCLASS code and the more complicated, but exact, hi\_CLASS code, is roughly $\sim 0.1\%$ accuracy across all multipoles, as shown on the right panel of Fig.~\ref{fig:classcls}.

Finally, we should also stress that an open point of debate in the community is the proper application of the quasi-static approximation, especially in more complicated models, as is extensively discussed in Ref.~\cite{Pace:2020qpj}. Even though this issue does not affect our effective fluid approach, it is a point that should be addressed prior to the advent of the next-generation surveys, in order to avoid unwanted theoretical errors in the predictions. Still, as demonstrated in this work, the effective fluid approach can provide a powerful framework, covering most viable MG models and allowing for a simple (and educational) way to modify the Boltzmann codes, which are necessary in data analyses.



\vspace{6pt} 




\funding{SN acknowledges support from the research project PID2021-123012NB-C43, and the Spanish Research Agency (Agencia Estatal de Investigaci\'on) through the Grant IFT Centro de Excelencia Severo Ochoa No CEX2020-001007-S, 
funded by MCIN/AEI/10.13039/501100011033.}

\institutionalreview{Not applicable}

\informedconsent{Not applicable}



\dataavailability{Not applicable.}

\acknowledgments{In this section you can acknowledge any support given which is not covered by the author contribution or funding sections. This may include administrative and technical support, or donations in kind (e.g., materials used for experiments).}

\conflictsofinterest{The authors declare no conflict of interest.}




\abbreviations{Abbreviations}{
The following abbreviations are used in this manuscript:\\

\noindent 
\begin{tabular}{@{}ll}
BBKS & Bardeen, Bond, Kaiser and Szalay (transfer function) \\
CAMB & Code for Anisotropies in the Microwave Background\\ 
CDM & Cold Dark Matter \\  
CLASS & Cosmic Linear Anisotropy Solving System \\
CMB & Cosmic Microwave Background\\
DE  & Dark Energy \\
DM & Dark Matter \\
EFCLASS & Effective Fluid CLASS\\
EOS & Equation of State \\
FLRW & Friedmann–Lema\^itre–Robertson–Walker metric\\  
GC & Galaxy Counts\\
GR & General Relativity \\
GW  & Gravitational Wave \\
ISW & Integrated Sachs-Wolfe effect \\
HDES & Horndeski Designer model\\
HS & Hu-Sawicki model\\
KGB & Kinetic Gravity Braiding model \\
\lcdm & The cosmological constant ($\Lambda$) and cold dark matter (CDM) model \\
LSS & Large Scale Structure \\ 
MG & Modified Gravity\\
SVT & Scalar-Vector-Tensor\\ 
\end{tabular}
}

\appendixtitles{no} 



\begin{adjustwidth}{-\extralength}{0cm}

\reftitle{References}


\bibliography{review}




%


\end{adjustwidth}
\end{document}